\documentclass[preprint,12pt]{elsarticle}

\usepackage{amsmath,amssymb,amsthm,mathtools}
\usepackage{booktabs}
\usepackage{microtype}
\usepackage[hidelinks]{hyperref}
\usepackage{graphicx}
\usepackage{tabularx}
\usepackage{longtable}

\makeatletter
\def\ps@pprintTitle{%
  \let\@oddhead\@empty
  \let\@evenhead\@empty
  \let\@oddfoot\@empty
  \let\@evenfoot\@empty
}
\makeatother

\newtheorem{proposition}{Proposition}
\newtheorem{corollary}{Corollary}

\newcommand{\BPi}{\boldsymbol{\Pi}}
\newcommand{\Bsigma}{\boldsymbol{\Sigma}}

\newcommand{\Bmu}{\boldsymbol{\mu}}

\newcommand{\Bbeta}{\boldsymbol{\beta}}

\newcommand{\Blambda}{\boldsymbol{\lambda}}
\newcommand{\transp}{\top}
\newcommand{\Tdagger}{\mathbf{T}_{\dagger}}
\newcommand{\Gam}{\operatorname{Gamma}}
\newcommand{\Bern}{\operatorname{Bernoulli}}
\newcommand{\Beta}{\operatorname{Beta}}
\biboptions{sort&compress}
\begin{document}
\begin{frontmatter}
\title{A Robust BGe Score for Gaussian Bayesian Network Structure Learning via Scale Mixtures}
\author[inst1]{Marco Grzegorczyk}
\ead{m.a.grzegorczyk@rug.nl}
\affiliation[inst1]{organization={Bernoulli Institute, University of Groningen},
  city={Groningen},country={Netherlands}}
 
\begin{abstract}
The Bayesian Gaussian equivalent (BGe) score is a widely used
metric for Gaussian Bayesian network structure learning because of its analytical tractability and score equivalence.
However, the Gaussian assumption makes the BGe score
sensitive to atypical observations (outliers).
We develop a robust framework based on Gaussian scale mixtures,
in which each multivariate observation has a latent precision
weight shared across its variables.
Conditional on these weights, the Gaussian likelihood and
Normal--Wishart conjugacy are retained.
We derive the corresponding weighted BGe score, which has
closed-form local marginal likelihoods, and show that score
equivalence is preserved even after integrating out the latent
precision weights under a graph-independent mixing prior.
Two familiar robust distributions are special cases:
Gamma mixing of observation precisions yields the multivariate
Student-$t$ distribution, whereas two-point mixing yields
a contaminated Gaussian distribution.
Although these two models differ in their treatment of atypical
observations, they share the same conditional network score.
Network structures and latent observation weights can thus be
inferred jointly within a unified Bayesian framework. \\
Our simulation studies demonstrate improved structure learning
accuracy in the presence of contamination, even under
misspecification of the contamination mechanism.
Applications to five real-world datasets illustrate the outlier
detection capabilities of the proposed models and the impact
of robust modeling on inferred network structures.
\end{abstract}
\begin{keyword}
Gaussian Bayesian networks \sep BGe score \sep
score equivalence   \sep robust structure learning  \sep Gaussian scale mixtures  
\end{keyword}
\end{frontmatter}

\section{Introduction}
\label{sec:introduction}
Gaussian Bayesian networks represent multivariate Gaussian
distributions through directed acyclic graphs (DAGs)
\citep{KollerFriedman2009,ScutariDenis2014}.
The DAGs yield factorizations of the joint distribution
which are useful for describing conditional independence
relations and for learning dependence structures from data
\citep{Chickering2002,TsamardinosEtAl2006}.
A well-established Bayesian scoring metric is the Bayesian
Gaussian equivalent (BGe) score, whose Normal--Wishart
construction yields closed-form local marginal likelihoods
and assigns equal scores to Markov-equivalent DAGs
\citep{GeigerHeckerman2002,KuipersEtAl2014}.\\
Learning the structure of Bayesian networks has received considerable
attention over the past decades, resulting in a wide range
of constraint-based, score-based, and hybrid learning algorithms \citep{Kitson2023},
many of which are implemented in the R package \texttt{bnlearn}
\citep{Scutari2010}.
In parallel, substantial methodological advances have been made
in Bayesian structure learning, particularly in the development
of efficient Markov chain Monte Carlo (MCMC) algorithms for sampling from the posterior
distribution over network structures
\citep{FriedKollerOrderMCMC,GrzegorczykHusmeierMach08,Kuipers_2017,KuipersEtAl2022,nikzad2025scalingbayesiandagsampling}.
While these developments have considerably improved the
computational feasibility of network structure learning,
the choice of an appropriate sampling distribution for the
observed data remains an important statistical issue.\\
The present paper continues a line of research aimed at extending Bayesian structure learning for Gaussian Bayesian networks (GBNs) beyond the classical setting. Earlier contributions addressed incomplete \citep{SCUTARI_REASONING,Gryze2023}, ordinal \citep{OSEM,Gryze2024}, and hybrid \citep{SCUTARI_NEERLANDICA,Gryze2025} data. The BGe framework was also extended to mixture GBNs \citep{GrzegorczykHusmeierBioinf08,grzegorczyk_TIED}, hierarchically coupled GBNs for multi-group data \citep{Scutari2022,PGM2026}, and dynamic GBNs with both intra- and inter-slice edges \citep{SOFTWARE3,IJAR_Kezhuo}. In this paper, we extend the BGe metric to robust structure learning in the presence of atypical observations. \\
The standard BGe score assumes a multivariate Gaussian distribution for every observation. Atypical observations can have a substantial effect on the sample covariance matrix and consequently on the score for competing network structures. Removing suspicious observations in a preprocessing step ignores uncertainty about whether these observations are actually outliers. Moreover, it separates outlier detection from network structure learning, although the two tasks may benefit from being performed jointly.
An alternative is to represent heavy tails or contamination by a mixture of Gaussian distributions with different scales. Gaussian scale mixtures form a classical robust modeling device \citep{AndrewsMallows1974}. In particular, Gamma mixing of Gaussian precisions yields a Student-$t$ distribution, often
used for robust estimation \citep{LangeLittleTaylor1989}. A two-point mixing distribution instead yields a contaminated Gaussian model, with one regular and one variance-inflated component \citep{PunzoMcNicholas2016}.\\
Gaussian scale mixtures have previously been employed
for robust undirected graphical model learning
\citep{FinegoldDrton2011,FinegoldDrton2014}
and for robust chain graph modeling
\citep{ChakrabortyEtAl2024}.
These approaches demonstrate how latent scale parameters
can be used to account for departures from Gaussianity
in graphical models.
Alternative approaches to robust and non-Gaussian DAG
structure learning include robustification of the PC
algorithm \cite{KALISCH_ROBUST} and rank-based conditional
independence testing \cite{HARRISDRton2013}.
More recently, \citet{NazariEtAl2026} combined Gaussian
scale mixtures with Bayesian DAG structure learning using
horseshoe shrinkage priors on the modified Cholesky factor
of the precision matrix. In contrast to their shrinkage-based approach, we focus on extending the BGe score while explicitly preserving its score equivalence property. \\
The proposed extension allows for robust Bayesian
model averaging over DAGs. By analytically integrating out the Gaussian network
parameters conditional on the latent precision weights,
we obtain a locally factorizing marginal likelihood
that can be used directly for MCMC sampling over DAGs.
An important property of our weighted BGe score is that
score equivalence is preserved not only for fixed precision
weights but also after integrating out these weights
under a DAG-independent mixing prior.\\
The proposed framework combines both continuous and
binary scale mixing for Bayesian network structure learning.
We introduce one positive precision weight for each
multivariate observation. Conditional on the weights,
the model is Gaussian, and the marginal likelihoods can
be derived analytically under a Normal--Wishart prior.
The resulting analytical expressions are the same
regardless of whether the weights follow a continuous
or discrete mixing distribution.
Rather than estimating a robust covariance matrix
before performing network structure learning,
our approach incorporates robustness directly
into Bayesian network inference.
The network structure and the latent precision
weights are inferred jointly, so that uncertainty
about the precision weights is propagated into
the posterior distribution over network structures.
In turn, inference on the observation-specific precision weights depends on the network structure. This provides a coherent Bayesian framework for robust structure learning and outlier detection.\\
The theoretical contributions of this paper are threefold.
First, we derive a weighted BGe score for arbitrary positive
observation weights and show how it relates to the standard
BGe score.
Second, we prove score equivalence both for fixed weights
and after integrating out the weights under a DAG-independent
mixing distribution.
Third, we demonstrate that, despite preserving score
equivalence, integrating out the weights does not necessarily
preserve the conditional independencies implied by the
underlying Gaussian DAG.\\
In our simulation study, we compare the structure learning
accuracy of the proposed robust models under different
contamination mechanisms and for different sample sizes.
We also apply the models to five real-world datasets
to evaluate their outlier detection capabilities and
investigate the impact of robust modeling on posterior
network inference.\\
The remainder of the paper is organized as follows:
Section~2 introduces the methodological framework,
derives the weighted BGe score, proves its score equivalence,
and describes the posterior inference algorithms.
Section~3 presents the simulation study.
Section~4 presents the applications to real-world data.
Section~5 concludes with a discussion.
Additional material is provided in the appendices.

\section{Methods}

\subsection{Gaussian scale mixture models for Bayesian networks}
\label{sec:framework}
Bayesian networks (BNs) use directed acyclic graphs (DAGs) to represent
conditional independencies among random variables. We consider $n$ variables
$X_1,\ldots,X_n$ and denote their DAG by $\mathcal{G}$. The parent set of
node $X_j$ is denoted by $\BPi_j$. Given $\mathcal{G}$, the joint density
factorizes according to the implied parent sets:
\begin{equation}\label{eq:bn-factor}
 p(X_1,\ldots,X_n\mid\mathcal{G})
 =\prod_{j=1}^n p(X_j\mid\BPi_j).
\end{equation}

The random variables can be stacked into
$\mathbf{X}=(X_1,\ldots,X_n)^{\transp}$. 
We observe a dataset
$\mathcal{D}=\{\mathbf{x}_1,\ldots,\mathbf{x}_m\}$ of $m$
realizations of $\mathbf{X}$, with $\mathbf{x}_i\in\mathbb{R}^n$.
For a Gaussian Bayesian network (GBN), the joint distribution is
$\mathcal{N}_n(\Bmu,\Bsigma)$, where $\Bsigma$ is consistent with the
conditional independencies implied by $\mathcal{G}$.

To make this framework robust against atypical observations, we introduce a
positive latent precision parameter $\lambda_i$ for each complete
observation $\mathbf{x}_i$. Conditional on $\lambda_i$, we assume
\begin{equation}\label{eq:weighted-normal}
 \mathbf{x}_i\mid(\Bmu,\Bsigma,\lambda_i,\mathcal{G})
 \sim \mathcal{N}_n\!\left(\Bmu,\frac{\Bsigma}{\lambda_i}\right),
 \qquad i=1,\ldots,m.
\end{equation}
Let $\Blambda=(\lambda_1,\ldots,\lambda_m)^{\transp}$ denote the
vector of observation-specific precision weights.
Conditional on $\Bmu$, $\Bsigma$, $\Blambda$ and
$\mathcal{G}$, the observations are assumed to be independent.
Thus $\lambda_i=1$ recovers the ordinary Gaussian distribution, whereas
$\lambda_i<1$ inflates the covariance matrix and reduces the influence
of observation $i$. The precision matrix becomes $\lambda_i\mathbf{W}$,
where $\mathbf{W}=\Bsigma^{-1}$. Since all variables within an observation share the same precision weight, downweighting takes place at the observation level rather than at the level of individual variable values.

Let $\psi$ denote the parameters of the mixing distribution. Conditional on $\psi$, the weights are assumed to be mutually independent, with a distribution that does not depend on the Gaussian parameters or the DAG:
\begin{equation}
\label{eq:mixing-distribution}
p(\Blambda\mid\psi,\Bmu,\Bsigma,\mathcal{G})
=
p(\Blambda\mid\psi)
=
\prod_{i=1}^m p(\lambda_i\mid\psi).
\end{equation}
We further assume that the mixing parameters $\psi$
are a priori independent of the Gaussian parameters
and the DAG, such that
$p(\psi\mid\Bmu,\Bsigma,\mathcal{G})=p(\psi)$.
For given $\Bmu$ and $\Bsigma$, integrating out
the precision weight yields
\begin{equation}
\label{eq:gaussian-scale-mixture}
p(\mathbf{x}_i\mid\Bmu,\Bsigma,\psi)
=
\int_0^\infty
\mathcal{N}_n\!\left(
\mathbf{x}_i\mid\Bmu,
\frac{\Bsigma}{\lambda_i}
\right)
p(\lambda_i\mid\psi)\,d\lambda_i.
\end{equation}

For continuous mixing distributions, $p(\lambda_i\mid\psi)$
denotes a probability density. For discrete mixing
distributions, it denotes a probability mass function,
and the integral is replaced by a sum.
We focus on two special cases of this
general framework: the continuous robust model (CRM),
based on Gamma-distributed precision weights, and the
binary robust model (BRM), based on a two-point mixing
distribution. Both models are compared with the standard
Gaussian Bayesian network model (BGe), for which
$\lambda_i=1$ for all $i$.

\subsection{Weighted BGe score}
\label{sec:bge}
In this section, we derive the weighted BGe score conditional on the precision weights and establish its score equivalence.

\subsubsection{Weighted sufficient statistics}
For a complete DAG $\mathcal{G}_C$, the likelihood from
Eq.~\eqref{eq:weighted-normal} can be written as
\begin{align}\label{eq:weighted-likelihood}
 p(\mathcal{D}\mid\Bmu,\mathbf{W},\Blambda,\mathcal{G}_C)
 &=(2\pi)^{-nm/2}|\mathbf{W}|^{m/2}
 \prod_{i=1}^m\lambda_i^{n/2}\\
 &\quad\times\exp\!\left\{-\frac12\sum_{i=1}^m\lambda_i
 (\mathbf{x}_i-\Bmu)^{\transp}\mathbf{W}
 (\mathbf{x}_i-\Bmu)\right\}.\nonumber
\end{align}
Define the sum of the weights, the weighted sample mean, and the
weighted scatter matrix by
\begin{equation}
 m_\lambda:=\sum_{i=1}^m\lambda_i,
 \qquad
 \bar{\mathbf{x}}_{[\lambda]}
 :=\frac{1}{m_\lambda}\sum_{i=1}^m\lambda_i\mathbf{x}_i,
 \qquad
 \mathbf{S}_{[\lambda]}
 :=\sum_{i=1}^m\lambda_i
 (\mathbf{x}_i-\bar{\mathbf{x}}_{[\lambda]})
 (\mathbf{x}_i-\bar{\mathbf{x}}_{[\lambda]})^{\transp}.
 \label{eq:weighted-statistics}
\end{equation}
The weighted sum of squares can be decomposed as
\begin{align}\label{eq:weighted-decomposition}
 \sum_{i=1}^m\lambda_i
 (\mathbf{x}_i-\Bmu)(\mathbf{x}_i-\Bmu)^{\transp}
 =\mathbf{S}_{[\lambda]}+m_\lambda
 (\bar{\mathbf{x}}_{[\lambda]}-\Bmu)
 (\bar{\mathbf{x}}_{[\lambda]}-\Bmu)^{\transp}.
\end{align}
Hence the Gaussian quadratic form, and therefore Normal--Wishart
conjugacy, are preserved conditional on $\Blambda$.

\subsubsection{Normal--Wishart prior and weighted marginal likelihood}
We assign the fully conjugate Normal--Wishart prior to the
mean vector and the precision matrix:
\begin{equation}\label{eq:normal-wishart}
 \mathbf{W}\mid\mathcal{G}_C\sim
 \mathcal{W}_n(\alpha_w,\Tdagger),\qquad
 \Bmu\mid(\mathbf{W},\mathcal{G}_C)\sim
 \mathcal{N}_n\!\left(\Bmu_0,
 (\alpha_\mu\mathbf{W})^{-1}\right),
\end{equation}
where $\alpha_w>n-1$, $\alpha_\mu>0$, and $\Tdagger$ is positive
definite. We use the \emph{rate} parameterization of the Wishart
distribution, so that its density contains the kernel
\begin{equation}\label{eq:wishart-kernel}
 p(\mathbf{W})\propto
 |\mathbf{W}|^{(\alpha_w-n-1)/2}
 \exp\!\left\{-\frac12\operatorname{tr}
 (\Tdagger\mathbf{W})\right\}.
\end{equation}
If $\alpha_w>n+1$, we can write
$\Tdagger=(\alpha_w-n-1)\mathbf{T}_0$, where
$\mathbf{T}_0=\mathbb{E}[\Bsigma]$ is the prior expectation of the Gaussian covariance matrix. For the commonly used default $\alpha_w=n+2$ and
$\mathbf{T}_0=\mathbf{I}_n$, we have $\Tdagger=\mathbf{I}_n$.
For any subset $L\subseteq\{X_1,\ldots,X_n\}$ with $l=|L|$ variables,
let $\mathcal{D}^{L}$ denote the corresponding subdataset. We write
$\Tdagger^{L,L}$, $\mathbf{S}_{[\lambda]}^{L,L}$,
$\bar{\mathbf{x}}_{[\lambda]}^{L}$ and $\Bmu_0^{L}$ for the
corresponding subvectors and submatrices.
The precision matrix of the marginal Gaussian
distribution for the variables in $L$ is
$\mathbf{W}_L=(\Bsigma^{L,L})^{-1}$, which in
general differs from the principal submatrix
$\mathbf{W}^{L,L}$ of the full precision matrix. Define
$\alpha_{w,L}:=\alpha_w-n+l$
with the updated rate matrix
 $$\mathbf{T}_{\ddagger}^{L,L}
 :={\Tdagger}^{L,L}+\mathbf{S}_{[\lambda]}^{L,L}
 \quad+\frac{\alpha_\mu m_\lambda}{\alpha_\mu+m_\lambda}
 (\bar{\mathbf{x}}_{[\lambda]}^{L}-\Bmu_0^{L})
 (\bar{\mathbf{x}}_{[\lambda]}^{L}-\Bmu_0^{L})^{\transp}.$$

\begin{proposition}[Weighted complete-DAG marginal likelihood]
\label{prop:subset}
For a complete DAG $\mathcal{G}_C^{L}$ and fixed positive
weights $\Blambda$, integration over $\mathbf{W}_L$
and $\Bmu^L$ under the induced Normal--Wishart prior gives
\begin{align}\label{eq:weighted-subset}
 &p(\mathcal{D}^{L}\mid\Blambda,\mathcal{G}_C^{L})
 \nonumber\\
 &=\pi^{-lm/2}
 \left(\prod_{i=1}^m\lambda_i\right)^{l/2}
 \left(\frac{\alpha_\mu}{\alpha_\mu+m_\lambda}\right)^{l/2}
 \frac{\Gamma_l((\alpha_{w,L}+m)/2)}
      {\Gamma_l(\alpha_{w,L}/2)}
 \frac{\det(\Tdagger^{L,L})^{\alpha_{w,L}/2}}
      {\det(\mathbf{T}_{\ddagger}^{L,L})^{(\alpha_{w,L}+m)/2}}.
\end{align}
Here $\Gamma_l$ denotes the multivariate Gamma function. For the
empty set we define $p(\mathcal{D}^{\varnothing}\mid\Blambda,\mathcal{G}_C^{\varnothing})=1$. The full derivation is given in \ref{app:weighted}.
\end{proposition}

It is important to distinguish between $m$ and $m_\lambda$. The precision scaling parameter of the Gaussian mean
is updated from $\alpha_\mu$ to $\alpha_\mu+m_\lambda$,
whereas the degrees of freedom of the Wishart
distribution are updated from $\alpha_{w,L}$
to $\alpha_{w,L}+m$.
In particular, the latter update depends on the actual sample size, not on the sum of the weights. No constraint $m_\lambda=m$ is required. For $\lambda_i=1$ ($i=1,\ldots,m$), we have
$m_\lambda=m$, and the weighted sample mean and
scatter matrix reduce to their unweighted counterparts.
Eq.~\eqref{eq:weighted-subset} then reduces to
the standard BGe subset marginal likelihood.

\subsubsection{Score factorization and score equivalence}
To extend the weighted marginal likelihood to arbitrary DAGs,
we adopt the assumptions underlying the classical BGe score
\cite{GeigerHeckGaussUAI, GeigerHeckerman2002}.
Specifically, we assume global parameter independence,
parameter modularity, and likelihood equivalence.
These assumptions are implemented by deriving the local
parameter priors from a common Normal--Wishart prior
specified for a complete DAG, as in
Eq.~\eqref{eq:normal-wishart}.
For every subset $L$ of $l$ variables, we use the
corresponding induced Normal--Wishart prior, with
Wishart degrees of freedom
$\alpha_{w,L}=\alpha_w-n+l$ and rate matrix
$\Tdagger^{L,L}$.
All hyperparameters are shared across DAGs.
The observation-specific precision weights $\Blambda$
are positive and common to all variables of the same
observation. Conditional on these weights, the local
Gaussian regression model for node $X_j$ can be written as
\begin{equation}\label{eq:weighted-local-regression}
 x_{ij}\mid(\mathbf{x}_{i,\BPi_j},\lambda_i)
 \sim\mathcal{N}\left(
 a_j+\Bbeta_j^{\transp}\mathbf{x}_{i,\BPi_j},
 \frac{\sigma_j^2}{\lambda_i}
 \right),
\end{equation}
where $\mathbf{x}_{i,\BPi_j}$ is the vector of observed values
of the parents of $X_j$ in observation $i$, $a_j$ is the
intercept, $\Bbeta_j$ is the vector of regression coefficients,
and $\sigma_j^2$ is the residual variance of the corresponding
unweighted Gaussian regression model.
The observation-specific scaling leaves the intercept
and regression coefficients unchanged and affects only
the conditional residual variance. Hence, conditional
on $\Blambda$, the Gaussian likelihood retains its
factorization according to the DAG, and the local
parameter priors can be constructed in the same way
as for the standard BGe score. Consequently, the
Geiger--Heckerman construction can be applied to the
weighted subset marginal likelihoods from
Eq.~\eqref{eq:weighted-subset}.
For any DAG $\mathcal{G}$, we obtain
\begin{equation}
\label{eq:factorization}
p(\mathcal{D}\mid\mathcal{G},\Blambda)
=
\prod_{j=1}^n
\frac{
p(\mathcal{D}^{\{X_j\}\cup\BPi_j}\mid
\Blambda,\mathcal{G}_C^{\{X_j\}\cup\BPi_j})
}{
p(\mathcal{D}^{\BPi_j}\mid
\Blambda,\mathcal{G}_C^{\BPi_j})
}.
\end{equation}
Here, $\mathcal{G}_C^{L}$ denotes a complete DAG
on the variable subset $L$.
An important property of this construction is that
$p(\mathcal{D}^{L}\mid\Blambda,\mathcal{G}_C^{L})$ depends only on the variable subset
$L$, the observed data, and the common precision weights.
In particular, it does not depend on the orientation
of the edges in a complete DAG. \\
Two DAGs are Markov equivalent if they imply the same
conditional independence relations. This is equivalent
to having the same skeleton and the same unshielded
colliders (v-structures) \citep{VermaPearl1990}.
A score is said to be score equivalent if it assigns
equal scores to Markov-equivalent DAGs.

\begin{proposition}[Score equivalence]
\label{prop:score-equiv}
Under the prior assumptions stated above, the weighted BGe score is score equivalent for any fixed positive weight vector $\Blambda$.
That is, for any two Markov-equivalent DAGs
$\mathcal{G}_{\star}$ and $\mathcal{G}_{\bullet}$,
we have
\begin{equation}
\label{eq:score-equivalence}
p(\mathcal{D}\mid\mathcal{G}_{\star},\Blambda)
=
p(\mathcal{D}\mid\mathcal{G}_{\bullet},\Blambda).
\end{equation}
The proof is provided in \ref{app:equiv}.
\end{proposition}

\begin{corollary}[Integration over the precision weights]
\label{cor:integrated-equiv}
If the prior on the latent precision weights and their
hyperparameters is independent of the DAG, the fully integrated
marginal likelihood $p(\mathcal{D}\mid\mathcal{G})$ is also score equivalent.
\end{corollary}
\begin{proof}
The conditional marginal likelihoods are equal for every fixed
$\Blambda$. Hence, integrating these equal functions with respect to
the same DAG-independent mixing prior preserves equality.
\end{proof}

Despite preserving score equivalence, integration
over the latent precision weights generally does
not preserve the local score factorization because
the weights are shared across all variables.
Let $s_j(\Blambda)$ denote the local score of node
$X_j$ in Eq.~\eqref{eq:factorization}.
In general, the integral of a product is not equal to the product of the individual integrals:
\begin{equation}\label{eq:not-commute}
 \int\!\left\{\prod_j s_j(\Blambda)\right\}
 p(\Blambda)\,d\Blambda
 \ne
 \prod_j\int\!s_j(\Blambda)p(\Blambda)\,d\Blambda.
\end{equation}
Our MCMC algorithm explicitly includes the latent weights, so that the decomposable score is evaluated conditional on their current values. While factors that depend on the precision weights but not on the DAG cancel when comparing competing DAGs for fixed precision weights, they do not generally cancel when comparing different precision weight vectors. These factors must therefore be included when computing the Metropolis--Hastings acceptance probabilities for proposals that change the latent precision weights.

\subsection{Special cases of Gaussian scale mixtures}\label{sec:models}
The proposed framework allows for a wide range of mixing distributions $p(\lambda_i\mid\psi)$. The choice of this distribution determines the robust sampling model but does not affect the form of Eq.~\eqref{eq:weighted-subset}. In this paper, we focus on two special cases, one with continuous and one with discrete precision weights.

\subsubsection{Continuous robust model (CRM)}
\label{sec:student}
For the continuous model, we assume independently over observations
\begin{equation}\label{eq:gamma-weights}
 \lambda_i\mid\nu\sim
 \Gam\!\left(\frac\nu2,\frac\nu2\right),
 \qquad \nu>0,
\end{equation}
with the Gamma distribution parameterized by shape and \emph{rate}, so that $\mathbb E[\lambda_i]=1$.
It is well known that the multivariate Student-$t$ distribution
can be represented as a Gaussian scale mixture with Gamma-distributed
precision weights \citep{LangeLittleTaylor1989}.
Integrating out $\lambda_i$ in
Eqs.~\eqref{eq:weighted-normal} and \eqref{eq:gamma-weights}
yields the marginal density
\begin{equation}\label{eq:t-density}
 p(\mathbf{x}_i\mid\Bmu,\Bsigma,\nu)
 =\frac{\Gamma((\nu+n)/2)}
       {\Gamma(\nu/2)(\nu\pi)^{n/2}|\Bsigma|^{1/2}}
 \left(1+\frac{\delta_i}{\nu}\right)^{-(\nu+n)/2},
\end{equation}
where $\delta_i=(\mathbf{x}_i-\Bmu)^\transp\Bsigma^{-1}
          (\mathbf{x}_i-\Bmu)$.
Hence, $\mathbf{x}_i\mid\Bmu,\Bsigma,\nu\sim
t_{\nu,n}(\Bmu,\Bsigma)$, where $\Bsigma$ is the
Student-$t$ \emph{scale} matrix.
The latent-weight representation is particularly convenient for robust
inference. Conditional on the Gaussian parameters,
\begin{equation}\label{eq:t-weight-conditional}
 \lambda_i\mid (\mathbf{x}_i,\Bmu,\Bsigma,\nu)
 \sim\Gam\!\left(\frac{\nu+n}{2},
                       \frac{\nu+\delta_i}{2}\right),
\end{equation}
so that
\begin{equation}\label{eq:t-expected-weight}
 \mathbb E[\lambda_i\mid\mathbf{x}_i,\Bmu,\Bsigma,\nu]
 =\frac{\nu+n}{\nu+\delta_i}.
\end{equation}
The expected weight decreases as the squared Mahalanobis distance increases. Eq.~\eqref{eq:t-weight-conditional} holds for fixed $\Bmu$ and $\Bsigma$. Once these parameters have been integrated out, the full conditional distribution of $\lambda_i$ takes a different form.
Smaller $\nu$ gives heavier tails, while the Gaussian model
is recovered as $\nu\to\infty$. The parameter $\Bmu$ is
the location parameter of the multivariate Student-$t$
distribution and equals its expectation only for $\nu>1$.
For $\nu>2$, the marginal covariance matrix is
$\nu\Bsigma/(\nu-2)$, while $\Bsigma$ itself is the
scale matrix. Rather than fixing the degrees of freedom parameter $\nu$ a priori, we infer it from the data. We assign the hyperprior
\begin{equation}\label{eq:nu-prior}
 \nu\sim\Gam(a_\nu,b_\nu),
\end{equation}
where $a_\nu$ and $b_\nu$ are the shape and rate parameters,
respectively. 

\subsubsection{Binary robust model (BRM)}
\label{sec:contaminated}
For the binary model, we introduce a contamination indicator
$z_i\in\{0,1\}$ and define
\begin{equation}\label{eq:binary-weights}
 z_i\mid p_\kappa\sim\Bern(p_\kappa),
 \qquad \lambda_i=\kappa^{-z_i},\qquad\kappa>1.
\end{equation}
Thus $\lambda_i=1$ for a regular observation and
$\lambda_i=1/\kappa$ for a contaminated observation.
Summing over the two possible values of $z_i$ in
Eqs.~\eqref{eq:weighted-normal} and \eqref{eq:binary-weights}
yields the marginal density
\begin{align}\label{eq:contaminated-mixture}
 p(\mathbf{x}_i\mid\Bmu,\Bsigma,p_\kappa,\kappa)
 =(1-p_\kappa)\cdot \mathcal{N}_n(\mathbf{x}_i\mid\Bmu,\Bsigma)
+p_\kappa\cdot \mathcal{N}_n(\mathbf{x}_i\mid\Bmu,\kappa\Bsigma).
\end{align}
This is the density of a contaminated Gaussian distribution
with a common mean vector and proportional covariance matrices.
The role of the squared Mahalanobis distance is explicit in the conditional
posterior odds:
\begin{equation}\label{eq:binary-odds}
 \frac{p(z_i=1\mid\mathbf{x}_i,\Bmu,\Bsigma,p_\kappa,\kappa)}
 {p(z_i=0\mid\mathbf{x}_i,\Bmu,\Bsigma,p_\kappa,\kappa)}
 =\frac{p_\kappa}{1-p_\kappa}\,
 \kappa^{-n/2}
 \exp\!\left\{\frac12\left(1-\frac1\kappa\right)\delta_i\right\}.
\end{equation}
Thus the posterior odds of contamination increase with $\delta_i$.
As in Eq.~\eqref{eq:t-weight-conditional}, this expression is
conditional on the Gaussian parameters rather than based on the collapsed
BGe posterior.
For $p_\kappa$, we assign
the prior
\begin{equation}\label{eq:beta-prior}
 p_\kappa\sim\Beta(a_\kappa,b_\kappa).
\end{equation}
Writing $n_1=\sum_i z_i$ and $n_0=m-n_1$, integration over
$p_\kappa$ gives
\begin{equation}\label{eq:beta-binomial}
 p(\mathbf{z})
 =\frac{B(a_\kappa+n_1,b_\kappa+n_0)}
        {B(a_\kappa,b_\kappa)},
\end{equation}
where $B$ is the beta function. For the inflation factor $\kappa$, we specify the prior
\begin{equation}\label{eq:kappa-prior}
\eta=\log(\kappa-1),\qquad
\eta\sim\mathcal{N}(\mu_\eta,\sigma_\eta^2),
\qquad \kappa=1+e^\eta.
\end{equation} We assume that $p_\kappa$ and $\eta$ are a priori independent.
The standard Gaussian model is recovered in the
limits $p_\kappa \to 0$ or $\kappa \to 1$.
For fixed $\Bmu,\Bsigma$, the marginal covariance in the binary model
is $[(1-p_\kappa)+p_\kappa\kappa]\Bsigma$.

\subsection{Conditional independence and marginal mixing}\label{sec:ci}
For fixed $\lambda_i$, the covariance matrix in
Eq.~\eqref{eq:weighted-normal} is proportional to $\Bsigma$.
Scaling a Gaussian covariance matrix by a positive constant
does not change its conditional independence properties.
Thus, for fixed Gaussian parameters, the conditional
independencies implied by the DAG hold conditional
on the precision weights.
However, they are not
necessarily preserved when the weights are integrated out,
as the following example shows.\\
\noindent\textbf{Example.}
Let $n=2$, $\Bmu=\mathbf{0}$, $\Bsigma=\mathbf{I}_2$
and suppose the DAG has no edges. Conditional on $\lambda_i$,
the two variables are independent.
If $\lambda_i\sim\Gam(\nu/2,\nu/2)$ for a finite $\nu>0$,
their marginal joint distribution is a bivariate Student-$t$
distribution with identity scale matrix.
The conditional distribution of the second variable
given the first is then
\begin{equation}\label{eq:conditional-t}
 X_{i2}\mid (X_{i1}=x)
 \sim t_{\nu+1,1}\!\left(0,
                  \frac{\nu+x^2}{\nu+1}\right).
\end{equation}
Since the conditional scale depends on $x$, the two variables
are not marginally independent, although they are independent
conditional on $\lambda_i$.

This does not contradict Proposition~\ref{prop:score-equiv}
or Corollary~\ref{cor:integrated-equiv}.
Although Markov-equivalent Gaussian DAGs have equal marginal
likelihoods after integrating out the precision weights,
the resulting mixture distribution does not necessarily
satisfy the conditional independencies implied by the DAG.

\subsection{Posterior inference and network learning}
\label{sec:inference}
In this section, we derive the joint posterior distribution,
describe the MCMC algorithm, and explain how the MCMC samples
are used for network structure inference and outlier detection.
\subsubsection{Posterior distribution}

The Gaussian mean vector and precision matrix are analytically
integrated out, while the mixing variables and their parameters
are included in the MCMC sampling.
We write $\theta=(\Blambda,\nu)$ for the CRM and
$\theta=(\mathbf{z},\eta)$ for the BRM, where $p_\kappa$
has been integrated out according to
Eq.~\eqref{eq:beta-binomial}.
The joint posterior distribution is given by
\begin{equation}\label{eq:posterior}
 p(\mathcal{G},\theta\mid\mathcal{D})
 \propto
 p(\mathcal{D}\mid\mathcal{G},\Blambda(\theta)) \cdot
  p(\mathcal{G})\cdot p(\theta),
\end{equation}
where $\Blambda(\theta)$ is the precision weight vector
determined by $\theta$. The prior on the DAG is denoted
by $p(\mathcal{G})$, and the prior on $\theta$ is given by
\begin{equation}
 p(\theta)=
 \begin{cases}
 p(\nu)\displaystyle\prod_{i=1}^m p(\lambda_i\mid\nu),
 & \text{for the CRM},\\[3mm]
 p(\mathbf{z})\cdot p(\eta),
 & \text{for the BRM}.
 \end{cases}
\end{equation}

Score equivalence implies equal posterior probabilities
for Markov-equivalent DAGs if they have equal prior probabilities.

\subsubsection{MCMC algorithm}

We use a Metropolis-within-Gibbs algorithm to sample
from the posterior distribution in Eq.~\eqref{eq:posterior}.
For both robust models, each MCMC iteration consists
of three successive Metropolis--Hastings updates:
Steps 1, 2a, and 2b for the CRM, and Steps 1, 3a,
and 3b for the BRM.
For the standard BGe model, all precision weights
are fixed at one, and only Step 1 is performed.

\medskip
\noindent\textbf{Step 1: Network structure.}
Let $\mathcal{N}(\mathcal{G})$ denote the set of
neighboring DAGs that can be obtained from the current
DAG $\mathcal{G}$ by a single edge addition, deletion,
or reversal, subject to acyclicity and a maximum
number of parents per node (fan-in restriction).
We propose a new DAG $\mathcal{G}^{\star}$ uniformly
from $\mathcal{N}(\mathcal{G})$, so that the forward
and backward proposal probabilities are
\begin{equation}
 q(\mathcal{G}^{\star}\mid\mathcal{G})
 =\frac{1}{|\mathcal{N}(\mathcal{G})|},
 \qquad
 q(\mathcal{G}\mid\mathcal{G}^{\star})
 =\frac{1}{|\mathcal{N}(\mathcal{G}^{\star})|}.
\end{equation}
Keeping the precision weights $\Blambda$ fixed,
we accept the proposed DAG with the
Metropolis--Hastings acceptance probability
\begin{equation}\label{eq:mh-dag}
 A_{\mathcal{G}}
 =
 \min\left\{
 1,\,
 \frac{p(\mathcal{D}\mid\mathcal{G}^{\star},\Blambda)}
      {p(\mathcal{D}\mid\mathcal{G},\Blambda)}
 \frac{p(\mathcal{G}^{\star})}{p(\mathcal{G})}
 \frac{|\mathcal{N}(\mathcal{G})|}
      {|\mathcal{N}(\mathcal{G}^{\star})|}
 \right\}.
\end{equation}
Due to the local score factorization in
Eq.~\eqref{eq:factorization}, only the local
scores of nodes whose parent sets change need to be
recomputed: one local score for an edge addition
or deletion and two local scores for an edge reversal.

\medskip
\noindent\textbf{Step 2a: Continuous precision weights (CRM).}
Keeping $\mathcal{G}$ and $\nu$ fixed, we select an observation $i\in\{1,\ldots,m\}$ uniformly
at random and propose a new precision weight
$\lambda_i^{\star}$ according to
\begin{equation}\label{eq:lambda-proposal}
 \log(\lambda_i^{\star})
 =
 \log(\lambda_i)+\epsilon_\lambda,
 \qquad
 \epsilon_\lambda\sim\mathcal{N}(0,s_\lambda^2).
\end{equation}
All other precision weights remain unchanged.
Let $\Blambda^{\star}$ denote the proposed weight vector.
The Metropolis--Hastings acceptance probability is
\begin{equation}\label{eq:mh-lambda}
 A_\lambda
 =
 \min\left\{
 1,\,
 \frac{p(\mathcal{D}\mid\mathcal{G},\Blambda^{\star})}
      {p(\mathcal{D}\mid\mathcal{G},\Blambda)} \cdot
 \frac{p(\lambda_i^{\star}\mid\nu)}
      {p(\lambda_i\mid\nu)} \cdot
 \frac{\lambda_i^{\star}}{\lambda_i}
 \right\}.
\end{equation}
The last factor accounts for the asymmetry of the
proposal distribution on the original scale.
The Gamma prior ratio is given by
\begin{equation}\label{eq:lambda-prior-ratio}
 \frac{p(\lambda_i^{\star}\mid\nu)}
      {p(\lambda_i\mid\nu)}
 =
 \left(\frac{\lambda_i^{\star}}{\lambda_i}\right)^{\nu/2-1}
 \exp\left\{-\frac{\nu}{2}
 (\lambda_i^{\star}-\lambda_i)\right\}.
\end{equation}

\medskip
\noindent\textbf{Step 2b: Degrees of freedom (CRM).}
Conditional on the precision weights $\Blambda$,
the posterior distribution of $\nu$ satisfies
\begin{equation}\label{eq:nu-conditional}
 p(\nu\mid\Blambda,\mathcal{G},\mathcal{D})
 \propto
 p(\nu)\prod_{i=1}^m p(\lambda_i\mid\nu),
\end{equation}
where $p(\nu)$ is the Gamma hyperprior specified in
Eq.~\eqref{eq:nu-prior}.
We propose a new value $\nu^{\star}$ according to
\begin{equation}\label{eq:nu-proposal}
 \log(\nu^{\star})
 =
 \log(\nu)+\epsilon_\nu,
 \qquad
 \epsilon_\nu\sim\mathcal{N}(0,s_\nu^2).
\end{equation}
The Metropolis--Hastings acceptance probability is
\begin{equation}\label{eq:mh-nu}
 A_\nu
 =
 \min\left\{
 1,\,
 \frac{p(\nu^{\star})}{p(\nu)} \cdot 
 \prod_{i=1}^m
 \frac{p(\lambda_i\mid\nu^{\star})}
      {p(\lambda_i\mid\nu)} \cdot
 \frac{\nu^{\star}}{\nu}
 \right\}.
\end{equation}
As in Step 2a, the last factor accounts for the asymmetry
of the proposal distribution on the original scale. Since the precision weights remain unchanged, the weighted
BGe score cancels from the acceptance ratio.

\medskip
\noindent\textbf{Step 3a: Binary contamination indicators (BRM).}
Keeping $\mathcal{G}$ and $\eta$ fixed, we select  an observation $i\in\{1,\ldots,m\}$ uniformly
at random and propose to flip its contamination indicator:
\begin{equation}\label{eq:z-proposal}
 z_i^{\star}=1-z_i.
\end{equation}
All other contamination indicators remain unchanged.
Let $\mathbf{z}^{\star}$ denote the proposed indicator
vector and $\Blambda^{\star}$ the corresponding vector
of precision weights, obtained according to
Eq.~\eqref{eq:binary-weights}.
Since the proposal is symmetric, the
Metropolis--Hastings acceptance probability is
\begin{equation}\label{eq:mh-z}
 A_z
 =
 \min\left\{
 1,\,
 \frac{p(\mathcal{D}\mid\mathcal{G},\Blambda^{\star})}
      {p(\mathcal{D}\mid\mathcal{G},\Blambda)}\cdot 
 \frac{p(\mathbf{z}^{\star})}{p(\mathbf{z})}
 \right\},
\end{equation}
where the prior probabilities of $\mathbf{z}^{\star}$ and $\mathbf{z}$ are computed with
Eq.~\eqref{eq:beta-binomial}.

\medskip
\noindent\textbf{Step 3b: Variance inflation factor (BRM).}
Keeping the contamination indicators $\mathbf{z}$ fixed,
we propose a new value $\eta^{\star}$ according to
\begin{equation}\label{eq:eta-proposal}
 \eta^{\star}=\eta+\epsilon_\eta,
 \qquad
 \epsilon_\eta\sim\mathcal{N}(0,s_\eta^2).
\end{equation}
The proposed variance inflation factor is
$\kappa^{\star}=1+\exp(\eta^{\star})$,
which ensures that $\kappa^{\star}>1$.
The corresponding precision weights are
\begin{equation}
 \lambda_i^{\star}=(\kappa^{\star})^{-z_i},
 \qquad i=1,\ldots,m.
\end{equation}
Let $\Blambda^{\star}$ denote the proposed weight vector.
The Metropolis--Hastings acceptance probability is
\begin{equation}\label{eq:mh-eta}
 A_\eta
 =
 \min\left\{
 1,\,
 \frac{p(\mathcal{D}\mid\mathcal{G},\Blambda^{\star})}
      {p(\mathcal{D}\mid\mathcal{G},\Blambda)}\cdot 
 \frac{p(\eta^{\star})}{p(\eta)}
 \right\},
\end{equation}
where $p(\eta)$ is the Gaussian prior density specified
in Eq.~\eqref{eq:kappa-prior}.
Since both the prior and the symmetric proposal
distribution are defined on the $\eta$ scale,
no additional Hastings or Jacobian correction is required.\\

The proposal standard deviations and other MCMC settings
are specified in Sections~\ref{sec:experiments}
and~\ref{sec:real_data}.

\subsubsection{Posterior summaries}

The MCMC samples obtained after burn-in are used
for network structure inference and outlier detection.
Let $S$ denote the number of MCMC samples used
for posterior inference.

\medskip
\noindent\textbf{Network structure.}
To account for Markov equivalence, we convert each
sampled DAG into its completed partially directed
acyclic graph (CPDAG).
Let $\mathcal{C}^{(s)}$ denote the CPDAG corresponding
to the $s$-th MCMC sample.
Let $\mathbf{A}^{(s)}$ denote the binary adjacency matrix
of $\mathcal{C}^{(s)}$, with undirected edges represented
as bidirectional edges.
To quantify the posterior support for individual
network features \citep{FriedmanEtAl2000,FriedKollerOrderMCMC},
we estimate the posterior edge probability for each
ordered pair of distinct nodes $(X_j,X_k)$ as
\begin{equation}\label{eq:posterior-edge}
 \widehat{\pi}_{jk}
 =\frac{1}{S}\sum_{s=1}^S A_{jk}^{(s)},
 \qquad j\ne k.
\end{equation}
An undirected edge in a sampled CPDAG thus
contributes to both directions.
The posterior edge probabilities can be used to
compare competing structure learning models
and to extract networks by applying a probability
threshold. The resulting networks are descriptive posterior
summaries and are not necessarily valid CPDAGs.

\medskip
\noindent\textbf{Outlier detection.}
For the CRM, we estimate the posterior mean of each
observation-specific precision weight as
\begin{equation}\label{eq:posterior-lambda}
 \widehat{\lambda}_i
 =
 \frac{1}{S}\sum_{s=1}^S\lambda_i^{(s)}.
\end{equation}
Observations with smaller posterior mean precision
weights are considered more atypical.
For the BRM, we estimate the posterior contamination
probability of observation $i$ as
\begin{equation}\label{eq:posterior-contamination}
 \widehat{p}_i = \frac{1}{S}\sum_{s=1}^S z_i^{(s)}.
\end{equation}
Observations with larger posterior contamination
probabilities are considered more atypical.
Both models can thus be used to rank observations
according to their degree of atypicality.
The BRM also allows us to classify observations
as regular or contaminated by applying a threshold
to their posterior contamination probabilities,
e.g., $\widehat{p}_i\geq 0.5$.

\section{Simulation study}
\label{sec:experiments}
In our simulation study, we compare the standard Gaussian
Bayesian network model (BGe), the continuous robust model (CRM),
and the binary robust model (BRM) in terms of their network
structure learning accuracy.

\subsection{Experimental design}

\medskip\noindent\textbf{Common prior specifications.}
For all three models, we assign a uniform prior over the space of DAGs without imposing any additional
fan-in restriction.
For the Normal--Wishart prior specified in
Eq.~\eqref{eq:normal-wishart}, we set
\begin{equation}
 \alpha_w=n+2,\qquad
 \alpha_\mu=1,\qquad
 \Bmu_0=\mathbf{0},\qquad
 \Tdagger=\mathbf{I}_n.
\end{equation}
For the CRM, we set $a_\nu=2$ and $b_\nu=0.5$
in Eq.~\eqref{eq:nu-prior}, yielding a prior mean of $4$ for $\nu$.
For the BRM, we set $a_\kappa=b_\kappa=1$
in Eq.~\eqref{eq:beta-prior}, corresponding to a uniform
prior on $p_\kappa$, and use
$\eta\sim\mathcal{N}(\log 2,0.5^2)$ in
Eq.~\eqref{eq:kappa-prior}, yielding a prior median
of $3$ for $\kappa$.

\medskip\noindent\textbf{MCMC settings and initialization.}
For each simulated dataset and each of the three
competing models, we run the corresponding MCMC
algorithm for 500,000 iterations.
The first 250,000 iterations serve as burn-in,
and the MCMC state is stored every 500 iterations.
For all three models, the MCMC simulation is
initialized with an empty DAG.
For the CRM, we initialize all precision weights
at $\lambda_i=1$ and set $\nu=4$.
For the BRM, we initialize all contamination
indicators at $z_i=0$ and set $\kappa=3$.
Hence, both robust models are initialized with
all observations being treated as regular
Gaussian observations.
The standard deviations of the Gaussian random
walk proposals described in
Section~\ref{sec:inference} are set to $s_\lambda=0.3$, $s_\nu=0.15$, and $s_\eta=0.5$.
The posterior network summaries are computed
from the MCMC samples retained after burn-in.
To assess MCMC mixing and identify potential convergence
problems, we visually inspected trace plots of the network
scores, the number of edges, and the additional
model-specific parameters.

\medskip\noindent\textbf{Generation of simulated network data.}
We consider networks with $n=10$ nodes and an expected
number of 20 edges. For each experimental setting,
we generate 25 independent datasets, with sample sizes
$m\in\{50,100,200,400\}$.
To generate a random DAG, we fix the topological ordering
$X_1,\ldots,X_n$ and independently include each possible
edge $X_j\to X_k$ with $j<k$ with probability
\begin{equation}
 p_{\mathrm{edge}}=\frac{20}{n(n-1)/2}.
\end{equation}

For each node $X_j$ with $q_j=|\BPi_j|>0$ parents,
we independently sample $q_j+1$ coefficients
\begin{equation}
 b_{jr}=s_{jr}u_{jr},
 \qquad r=1,\ldots,q_j+1,
\end{equation}
where $u_{jr}\sim\mathcal{U}(1,2)$ and
$s_{jr}\in\{-1,+1\}$ with equal probabilities.
We normalize these coefficients as follows:
\begin{equation}
 \widetilde{b}_{jr}
 =
 \frac{b_{jr}}
 {\sqrt{\sum_{h=1}^{q_j+1}b_{jh}^2}}.
\end{equation}
The first $q_j$ normalized coefficients are assigned
to the parents of $X_j$ as regression coefficients
$\beta_{jk}$ ($X_k\in\BPi_j$).
The absolute value of the last coefficient determines
the residual standard deviation $\sigma_j$.
For nodes without parents, we set $\sigma_j=1$.
Hence, for every node,
\begin{equation}
 \sum_{X_k\in\BPi_j}\beta_{jk}^2+\sigma_j^2=1.
\end{equation}

The components of the mean vector $\Bmu$ are independently
sampled from a standard Gaussian distribution.
For given precision weights $\Blambda$, the observations
are generated recursively according to the topological
ordering of the DAG:
\begin{equation}\label{eq:simulation-generation}
 x_{ij}
 =
 \mu_j+
 \sum_{X_k\in\BPi_j}\beta_{jk}(x_{ik}-\mu_k)
 +
 \frac{\sigma_j}{\sqrt{\lambda_i}}\epsilon_{ij},
\end{equation}
where the $\epsilon_{ij}$ are mutually independent
standard Gaussian random variables.
This construction yields observations from the
Gaussian scale mixture model in
Eq.~\eqref{eq:weighted-normal}.\\

We consider three data-generating scenarios:
\begin{enumerate}
\item \emph{\textbf{Binary contamination}:}
The contamination indicators are independently sampled
from $\Bern(p_\kappa)$, with $p_\kappa=0.2$.
The precision weights are set to
$\lambda_i=\kappa^{-z_i}$, with
$\kappa\in\{1,4,9,16\}$.
The setting $\kappa=1$ corresponds to uncontaminated
Gaussian data.
\item \emph{\textbf{Continuous heavy tails}:}
The observation-specific precision weights $\lambda_i$
($i=1,\ldots,m$) are independently sampled from
$\Gam(\nu/2,\nu/2)$, with $\nu\in\{0.5,1,2,4,8\}$.
\item \emph{\textbf{Additive contamination (model mismatch)}:}
We first generate Gaussian network data by setting
$\lambda_i=1$ for all $i$ in
Eq.~\eqref{eq:simulation-generation}.
We then add Gaussian noise to a randomly selected
subset of observations:
\begin{equation}\label{eq:additive-contamination}
 \mathbf{y}_i=\mathbf{x}_i+z_i\mathbf{e}_i,
 \qquad
 z_i\sim\Bern(p_{\mathrm{add}}),
 \qquad
 \mathbf{e}_i\sim
 \mathcal{N}_n(\mathbf{0},\tau^2\mathbf{I}_n).
\end{equation}
The contamination indicators and noise vectors are
sampled independently of each other and of the
original observations.
We set $p_{\mathrm{add}}=0.2$ and consider
$\tau^2\in\{1,4,9,16\}$.
The contaminated observations $\mathbf{y}_i$ are
used for network structure learning.
\end{enumerate}
The first two scenarios correspond to the sampling
distributions underlying the BRM and the CRM,
respectively. In contrast, under additive contamination,
the covariance matrices of regular and contaminated
observations are
\begin{equation}
 \Bsigma_0=\Bsigma,\qquad
 \Bsigma_1=\Bsigma+\tau^2\mathbf{I}_n.
\end{equation}
Since these two covariance matrices are generally not
proportional, the additive contamination mechanism
cannot generally be represented by a Gaussian scale
mixture. This third scenario therefore allows us to
investigate the robustness of the CRM and the BRM under model misspecification. \\
In all three scenarios, the simulated data are used
directly for network structure learning without
prior standardization.

\medskip\noindent\textbf{Evaluation of structure learning accuracy.}
For each simulated dataset, we convert the true DAG
into its corresponding CPDAG. Let $\mathbf{C}^{\mathrm{true}}$
denote its $n\times n$ adjacency matrix, where an
undirected edge is represented by two entries equal
to one, one for each direction.
For each of the three competing models, we compute
the posterior edge probabilities $\widehat{\pi}_{jk}$
according to Eq.~\eqref{eq:posterior-edge}.
We evaluate these probabilities against the corresponding
binary entries of $\mathbf{C}^{\mathrm{true}}$.
All $n(n-1)$ off-diagonal entries are included in
the evaluation, while diagonal entries are excluded.
As a measure of structure learning accuracy, we use
the area under the precision--recall curve (AUPRC).
The AUPRC values are computed using the R package
\texttt{precrec} \citep{SaitoRehmsmeier2017}
and range between zero and one, with higher values
indicating better structure recovery. The baseline AUPRC for an uninformative classifier
corresponds to the proportion of positive entries
among the off-diagonal entries of the true CPDAG
adjacency matrix. For each experimental setting, the distribution of the AUPRC values obtained across the 25 independent repetitions is summarized by boxplots.

\begin{figure}[t!]
\centering
\includegraphics[width=\textwidth]{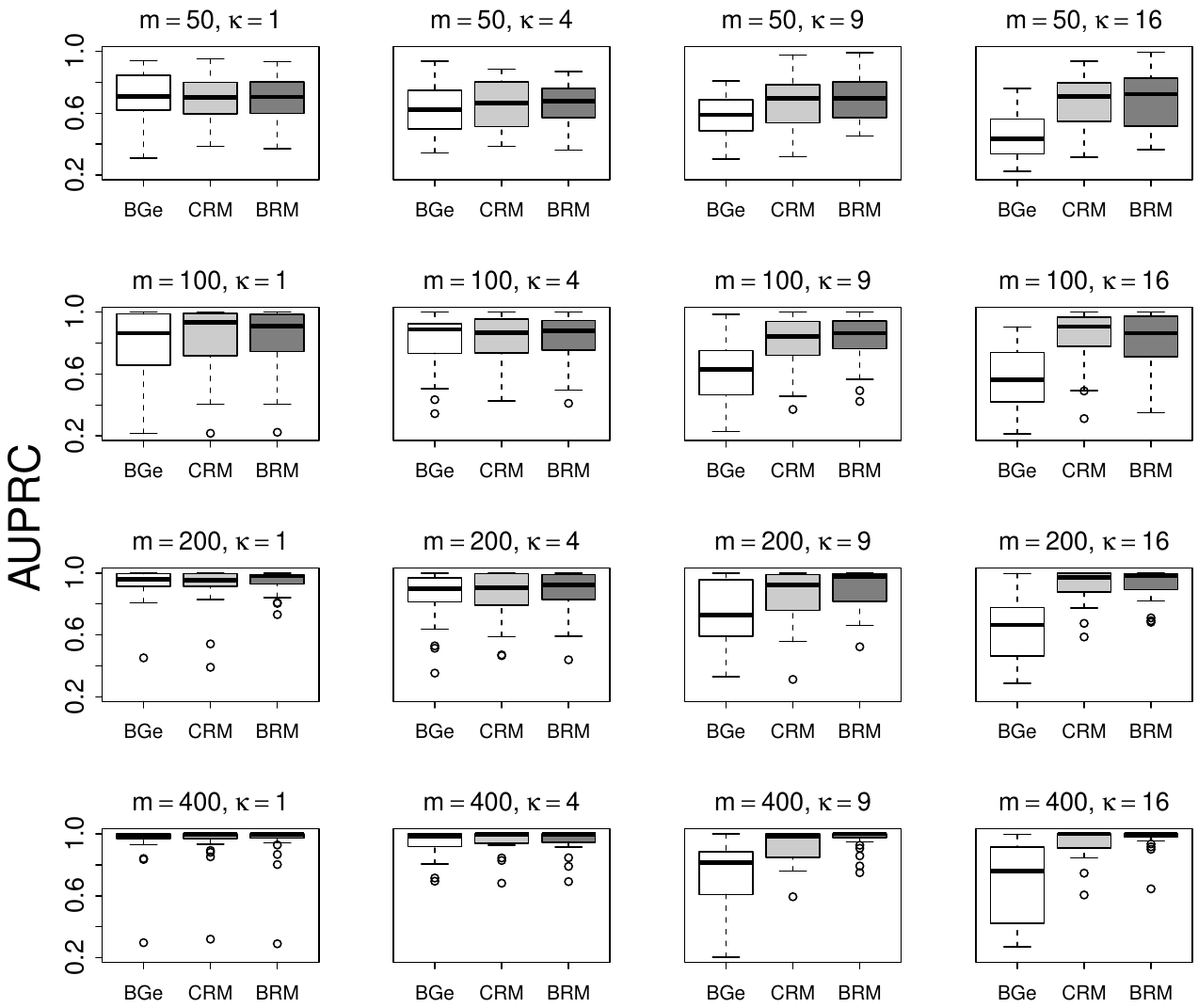}
\caption{\textbf{Binary contamination.}
Structure learning performance on data generated
under a binary Gaussian scale mixture, for which
the BRM is correctly specified when $\kappa>1$.
The contamination probability is fixed at $0.2$.
The rows correspond to sample sizes
$m\in\{50,100,200,400\}$, and the columns to
variance inflation factors $\kappa\in\{1,4,9,16\}$.
While $\kappa=1$ corresponds to uncontaminated
Gaussian data, larger values of $\kappa$ induce
increasingly severe contamination.
The boxplots summarize the AUPRC values across
25 independent simulation runs.
White, light gray, and dark gray represent
BGe, CRM, and BRM, respectively.
Higher AUPRC values indicate better structure recovery.}
\label{fig:binary_simulations}
\end{figure}

\begin{figure}[t!]
\centering
\includegraphics[width=\textwidth]{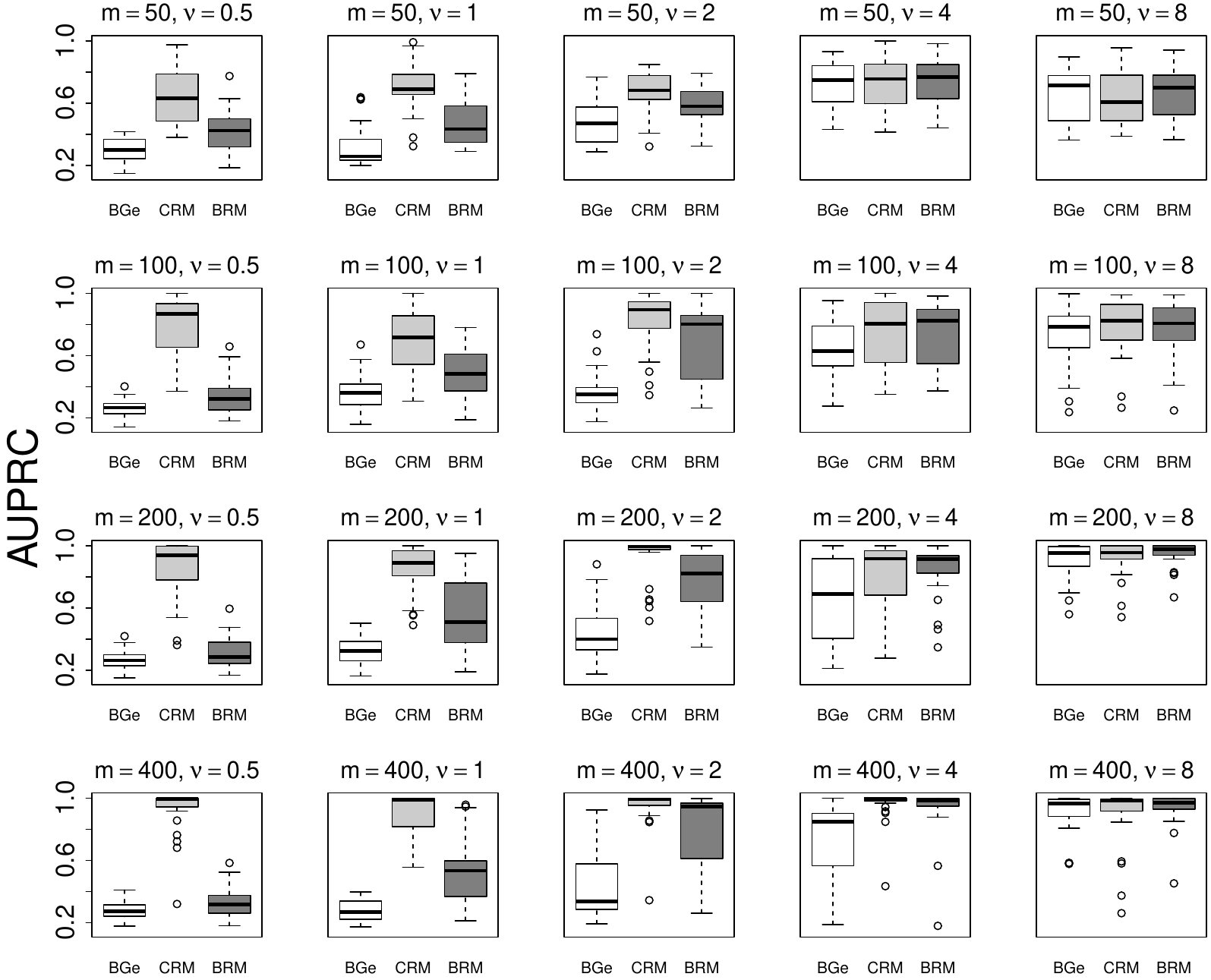}
\caption{\textbf{Continuous heavy tails.}
Structure learning performance on data generated
under a multivariate Student-$t$ distribution,
for which the CRM is correctly specified.
The columns correspond to degrees of freedom
$\nu\in\{0.5,1,2,4,8\}$, with smaller values
inducing heavier tails.
The sample sizes and boxplot conventions are
as in Figure~\ref{fig:binary_simulations}.}
\label{fig:continuous_simulations}
\end{figure}

\begin{figure}[t!]
\centering
\includegraphics[width=\textwidth]{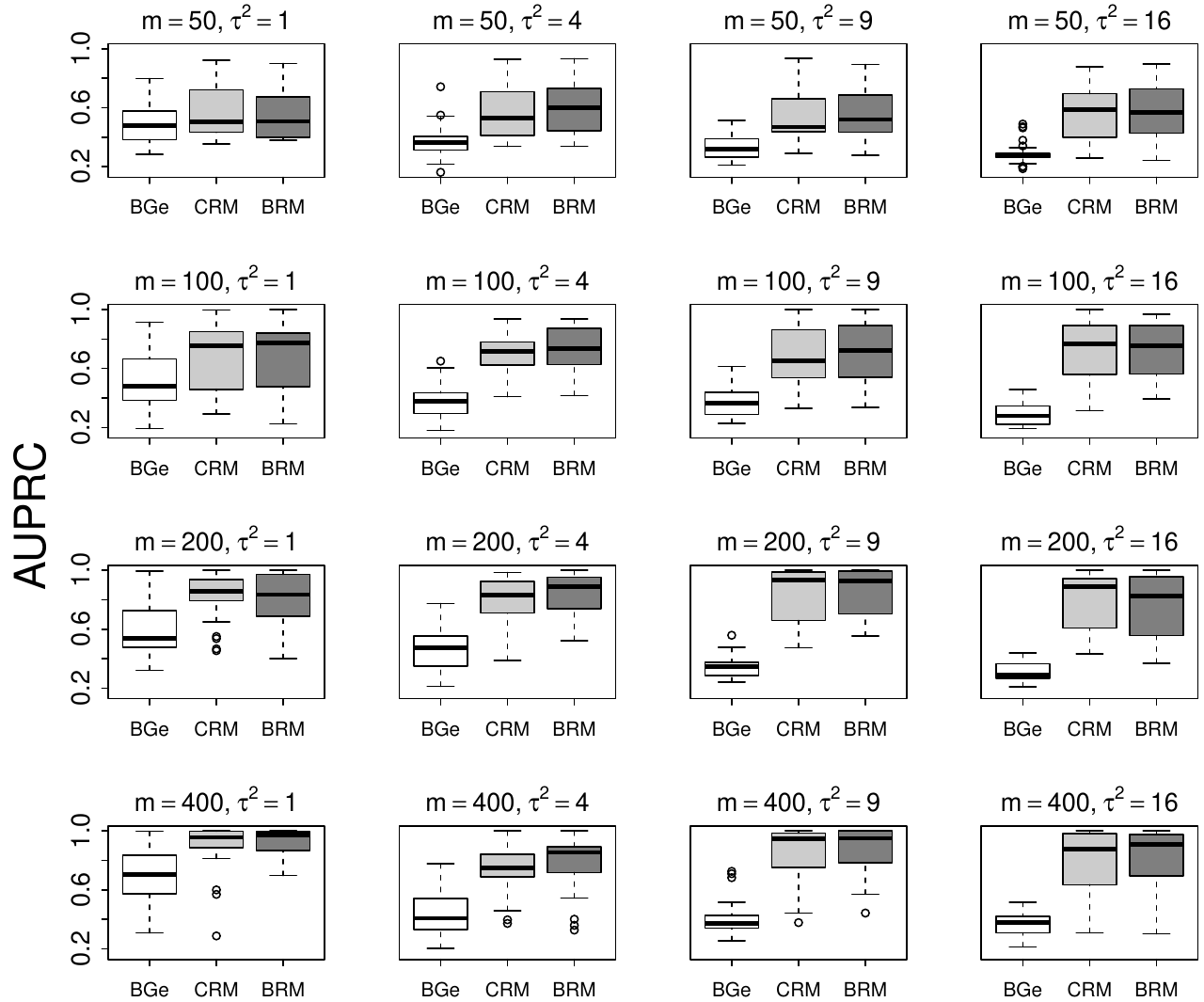}
\caption{\textbf{Additive contamination (model misspecification).}
Structure learning performance on Gaussian network
data contaminated by additive Gaussian noise,
for which neither robust model is generally
correctly specified.
The contamination probability is fixed at $0.2$.
The columns correspond to noise variances
$\tau^2\in\{1,4,9,16\}$, with larger values
inducing more severe contamination.
The sample sizes and boxplot conventions are
as in Figure~\ref{fig:binary_simulations}.}
\label{fig:additive_simulations}
\end{figure}

\subsection{Simulation results}
Figure~\ref{fig:binary_simulations} shows the structure
learning results for data generated under the binary
scale mixture model. For uncontaminated Gaussian data ($\kappa=1$), all three models perform comparably, suggesting that robust modeling does not come at the expense of substantially reduced structure learning accuracy.
For $\kappa=4$, the differences between the three models
remain relatively small. However, as the contamination strength increases ($\kappa=9$ and $\kappa=16$), both robust models
increasingly outperform the standard BGe model.
Although the BRM is the correctly specified model
for $\kappa>1$, the CRM achieves comparable structure
learning accuracy, with the BRM showing a slight
advantage in some experimental settings.
For the largest sample size ($m=400$), both robust
models achieve median AUPRC values close to one
across all contamination strengths, whereas the
performance of the standard BGe model remains
substantially impaired under strong contamination.
These results demonstrate the benefits of robust
structure learning and suggest that the CRM is
relatively insensitive to misspecification of the
underlying mixing distribution.\\
Figure~\ref{fig:continuous_simulations} shows the structure
learning results for data generated under the continuous
scale mixture model.
For extremely heavy tails ($\nu=0.5$ and $\nu=1$),
the correctly specified CRM substantially outperforms
both the standard BGe model and the BRM.
While the structure learning accuracy of the CRM
improves considerably with increasing sample size,
the performance of the standard BGe model remains poor.
The BRM also struggles to accommodate such heavy tails.
For $\nu=2$, both robust models outperform the standard
BGe model, but the CRM generally achieves higher
and less variable AUPRC values than the BRM.
For $\nu=4$, the differences between the two robust
models become smaller, and for $\nu=8$, all three
models perform comparably. \\
Figure~\ref{fig:additive_simulations} shows the structure
learning results under the additive contamination scenario.
Recall that neither the CRM nor the BRM is correctly
specified under this contamination mechanism.
Nevertheless, both robust models substantially outperform
the standard BGe model in most experimental settings.
The differences are particularly pronounced for larger
sample sizes and stronger contamination. While both robust models benefit considerably from increasing the sample size, the structure learning accuracy of the standard BGe model remains poor under strong contamination. Interestingly, the CRM and the BRM achieve very similar structure learning accuracies across the different experimental settings. These results demonstrate that the benefits of the proposed robust models are not restricted to data generated under their underlying Gaussian scale mixture assumptions. \\
Taken together, the results in Figures~\ref{fig:binary_simulations}, \ref{fig:continuous_simulations}, and
\ref{fig:additive_simulations} demonstrate that
the benefits of robust structure learning are not
restricted to correctly specified models. However, we observe an interesting asymmetry with respect to misspecification of the mixing distribution. While the CRM performs well on data generated under the binary scale mixture model, the BRM is considerably less successful on data generated under the continuous scale mixture model when the tails are extremely heavy. Under additive contamination, for which neither robust
model is correctly specified, both models achieve substantial and comparable improvements over the standard BGe model. Overall, these results suggest that continuous scale mixing offers greater flexibility in accounting for different degrees of atypicality among observations and may therefore be preferable when the underlying contamination mechanism is unknown.

\begin{table}[t!]
\centering
\caption{Characteristics of the real-world datasets
and literature-based outlier classifications.
Here, $m$ denotes the number of observations and $n$
the number of variables.}
\label{tab:real_datasets}
\small
\begin{tabularx}{\textwidth}{lccXl}
\toprule
\textbf{Dataset} & $m$ & $n$ & Outlier benchmark & References \\
\midrule
\textbf{CARS} & 111 & 11 &
Orthogonal: 103, 104, 107, 109, 111;\newline
Bad leverage (group 1): 25, 30, 32, 34, 36;\newline
Bad leverage (group 2): 102, 105, 106, 108, 110.
& \citep{Hubert2005} \\
\addlinespace
\textbf{BUSHFIRE} & 38 & 5 &
Clusters: 7--11 and 32--38;\newline
Additional: 12, 13, 31.
& \citep{MaronnaYohai1995,RockeWoodruff1996,
BeguinHulliger2008} \\
\addlinespace
\textbf{PULP FIBER} & 62 & 8 &
Primary: 51, 52, 56, 61;\newline
Additional: 22, 46--48, 58--60, 62.
& \citep{Rousseeuw2004,OngAlih2015} \\
\addlinespace
\textbf{MILK} & 86 & 8 &
Most extreme: 1, 2, 41, 70;\newline
Other outliers: 3, 12--17, 44, 47, 74, 75;\newline
Borderline: 11, 20, 27, 77.
& \citep{RockeWoodruff1996} \\
\addlinespace
\textbf{APPALACHIA} & 104 & 3 &
Classical and robust: 30, 90, 104;\newline
Robust only: 71, 99.
& \citep{Neykov2007} \\
\bottomrule
\end{tabularx}
\end{table}

\section{Real-world data applications}
\label{sec:real_data}

We apply the three competing models to five real-world
benchmark datasets from the robust statistics literature
that are suitable for Bayesian network structure learning.
Since the true network structures are unknown,
we focus on the outlier detection performance of the
two robust models and the impact of robust modeling
on the inferred network structures.

\subsection{Datasets and literature-based outlier benchmarks}

\begin{figure}[t!]
\centering
\includegraphics[width=\textwidth]
{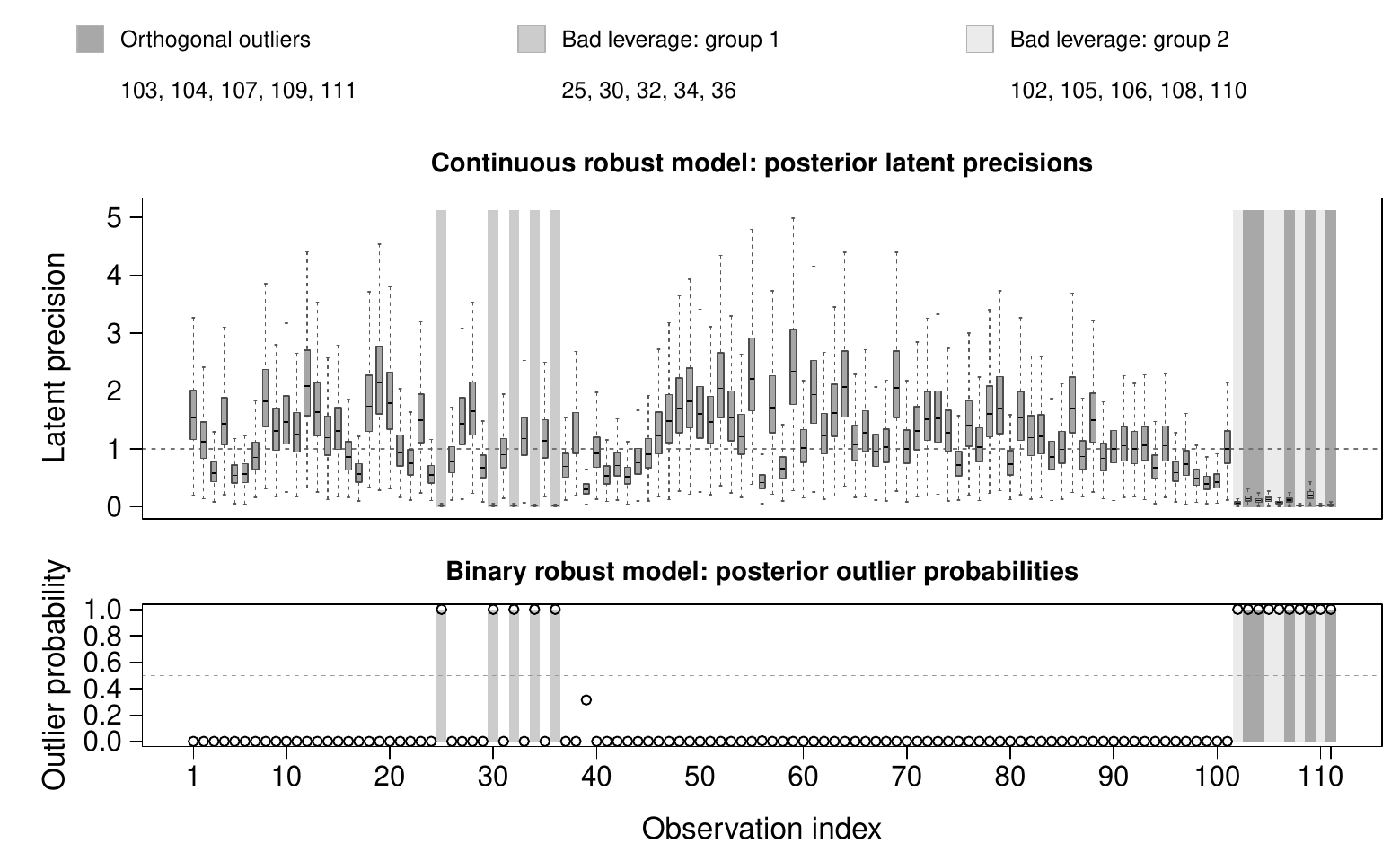}
\vspace{-0.85cm}
\caption{\textbf{Outlier detection in the CARS dataset.}
The upper panel shows the posterior distributions of the
observation-specific precision weights obtained with the CRM.
The lower panel shows the posterior contamination probabilities
obtained with the BRM.
Three shades of gray distinguish the groups of orthogonal
outliers and bad leverage points identified in the literature
\citep{Hubert2005}.}
\label{fig:cars_outliers}
\end{figure}

\begin{figure}[t!]
\centering
\includegraphics[width=\textwidth]{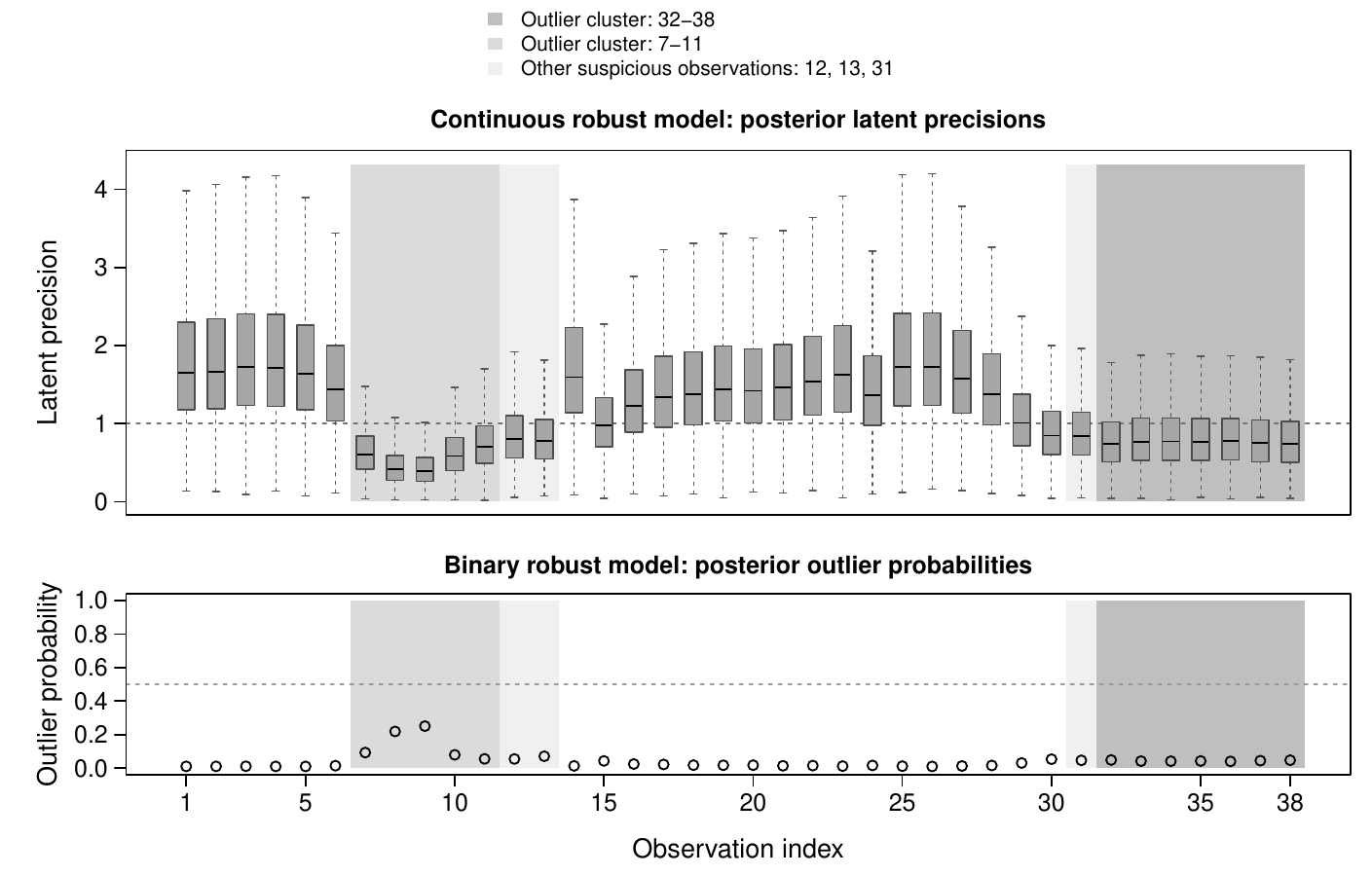}
\vspace{-0.75cm}
\caption{\textbf{Outlier detection in the BUSHFIRE dataset.}
Three shades of gray distinguish the two main outlier
clusters and the additional potentially atypical observations.
The visualization conventions are as in
Figure~\ref{fig:cars_outliers}.}
\label{fig:bushfire_outliers}
\end{figure}

Table~\ref{tab:real_datasets} summarizes the characteristics
of the real data and the outlier classifications reported in the robust statistics literature. Where appropriate, we distinguish between primary and expanded outlier benchmarks to account for different degrees of evidence for atypicality. The variable names and abbreviations used in the
network visualizations are listed in
\ref{app:variables}. \\ For the BUSHFIRE, PULP FIBER, MILK, and APPALACHIA
datasets, we consider both a primary and an expanded
outlier benchmark, comprising 12 and 15, 4 and 12,
15 and 19, and 3 and 5 observations, respectively.
 For the CARS dataset, we consider a single benchmark comprising the 15 orthogonal outliers and bad leverage points identified in the literature.  \\
The CARS dataset consists of 111 observations and 11 variables
describing different characteristics of cars.
The dataset was investigated by \citet{Hubert2005} using
robust principal component analysis.
Based on their diagnostic analysis, the authors identified
15 potentially outlying observations, which can be divided
into three distinct groups.
The first group comprises five orthogonal outliers
(103, 104, 107, 109, and 111).
The second group consists of five bad leverage points
(25, 30, 32, 34, and 36), while the third group contains
another five bad leverage points
(102, 105, 106, 108, and 110).
The authors also identified observations 6 and 96
as good leverage points, which are not included in our
outlier benchmark.
The three groups of potentially outlying observations
are highlighted using different shades of gray in
Figure~\ref{fig:cars_outliers}.\\
The BUSHFIRE dataset consists of 38 observations and five
variables representing satellite measurements used to
identify bushfire scars. The dataset has been extensively
studied in the literature on multivariate outlier detection
\citep{MaronnaYohai1995,RockeWoodruff1996,BeguinHulliger2008}.
In particular, \citet{MaronnaYohai1995} identified two
main groups of outliers: observations 7--11 and 32--38.
The second group is particularly challenging because
it can be masked by the first group when conventional
robust estimation methods are applied.
Subsequently, \citet{RockeWoodruff1996} identified
observation 31 as an additional outlier, while
\citet{BeguinHulliger2008} discussed observations
12 and 13 as potentially outlying observations.
The two main groups constitute our primary benchmark,
while the expanded benchmark additionally includes
observations 12, 13, and 31.
The three groups are highlighted using different
shades of gray in Figure~\ref{fig:bushfire_outliers}. \\

Detailed descriptions of the PULP FIBER, MILK,
and APPALACHIA datasets, together with the corresponding
outlier detection figures, are provided in
\ref{app:outlier_results}.

\subsection{Experimental settings and evaluation criteria}
Before applying the models, we standardize each variable
to have sample mean zero and sample variance one.
To allow more time for convergence and obtain more
accurate posterior estimates, we use substantially
longer MCMC runs than in the simulation study.
For each dataset and each model, we run the corresponding
MCMC algorithm for 5,000,000 iterations, with the first
2,500,000 iterations serving as burn-in.
The MCMC state is stored every 100 iterations.
All prior specifications, initializations, and proposal
standard deviations are identical to those used in the
simulation study in Section~\ref{sec:experiments}.
MCMC mixing was assessed as in
the simulation study.

\medskip\noindent\textbf{Evaluation of outlier detection.}
We evaluate outlier detection performance against
the primary and, where applicable, expanded
literature-based benchmarks summarized in
Table~\ref{tab:real_datasets}.
For the CRM, we rank the observations in ascending
order of their posterior mean precision weights
$\widehat{\lambda}_i$, whereas for the BRM we rank
the observations in descending order of their
posterior contamination probabilities
$\widehat{p}_i$.
Let $K$ denote the number of outliers in the
respective benchmark. We consider three measures
of outlier ranking performance:
\begin{itemize}
 \item \textbf{Top-}$\mathbf{K}$: the number of benchmark outliers
 among the $K$ highest-ranked observations.
 \item \textbf{Average precision} (\textbf{AP}): the weighted average
 of precision values along the precision--recall
 curve, with the increases in recall as weights.
 \item \textbf{Area under the receiver operating
 characteristic curve} (\textbf{AUC}): a measure of the
 ability to rank benchmark outliers above
 the remaining observations.
\end{itemize}
Higher values indicate better agreement with
the literature-based outlier classifications.
For the BRM, we additionally investigate the
classification obtained by applying a threshold
of 0.5 to the posterior contamination probabilities.
We report the total number of flagged observations,
the numbers of true positives (TP) and false positives (FP),
and the corresponding precision $\mathrm{TP}/(\mathrm{TP}+\mathrm{FP})$
and recall $\mathrm{TP}/K$.
Precision is undefined when no observations are flagged.
Since the literature classifications are
not necessarily exhaustive or unambiguous,
these  measures quantify agreement
with existing outlier detection results rather
than accuracy with respect to a known ground truth.

\medskip\noindent\textbf{Comparison of posterior network structures.}
To investigate the impact of robust modeling on network
inference, we compare the posterior edge probabilities
obtained with the three competing models.
For each pair of models, we compute the Pearson
correlation coefficient between the corresponding
posterior edge probabilities, excluding self-loops.

\begin{table}[t!]
\centering
\caption{\textbf{Outlier detection performance of the CRM and BRM.} Top-$K$ reports the number of
literature-based outliers recovered among the $K$ highest-ranked
observations. AP denotes average precision and AUC denotes
the area under the ROC curve.}
\label{tab:outlier_ranking}
\small
\setlength{\tabcolsep}{4pt}
\begin{tabular}{llrcccccc}
\toprule
& & &
\multicolumn{3}{c}{\textbf{CRM}} &
\multicolumn{3}{c}{\textbf{BRM}} \\
\cmidrule(lr){4-6}
\cmidrule(lr){7-9}
Dataset & Benchmark & $K$ &
Top-$K$ & AP & AUC &
Top-$K$ & AP & AUC \\
\midrule
CARS & Primary & 15 & 15 & 1.000 & 1.000 & 15 & 1.000 & 1.000 \\
\addlinespace
BUSHFIRE & Primary & 12 & 12 & 1.000 & 1.000 & 8 & 0.797 & 0.901 \\
 & Expanded & 15 & 15 & 1.000 & 1.000 & 13 & 0.940 & 0.965 \\
\addlinespace
PULP FIBER & Primary & 4 & 3 & 0.679 & 0.983 & 3 & 0.762 & 0.989 \\
 & Expanded & 12 & 11 & 0.965 & 0.992 & 10 & 0.950 & 0.987 \\
\addlinespace
MILK & Primary & 15 & 14 & 0.991 & 0.998 & 14 & 0.991 & 0.998 \\
 & Expanded & 19 & 17 & 0.982 & 0.994 & 17 & 0.951 & 0.978 \\
\addlinespace
APPALACHIA & Primary & 3 & 2 & 0.917 & 0.997 & 2 & 0.917 & 0.997 \\
 & Expanded & 5 & 5 & 1.000 & 1.000 & 5 & 1.000 & 1.000 \\
\bottomrule
\end{tabular}
\end{table}

\begin{table}[t!]
\centering
\caption{\textbf{Outlier detection results for the BRM using a posterior contamination probability threshold
of 0.5.} TP and FP denote the numbers of correctly and
incorrectly flagged observations relative to the published
benchmark classifications.}
\label{tab:outlier_threshold}
\small
\begin{tabular}{llrccccc}
\toprule
Dataset & Benchmark & $K$ &
Flagged & TP & FP & Precision & Recall \\
\midrule
CARS & Primary & 15 & 15 & 15 & 0 & 1.000 & 1.000 \\
\addlinespace
BUSHFIRE & Primary & 12 & 0 & 0 & 0 & -- & 0.000 \\
 & Expanded & 15 & 0 & 0 & 0 & -- & 0.000 \\
\addlinespace
PULP FIBER & Primary & 4 & 9 & 4 & 5 & 0.444 & 1.000 \\
 & Expanded & 12 & 9 & 8 & 1 & 0.889 & 0.667 \\
\addlinespace
MILK & Primary & 15 & 6 & 6 & 0 & 1.000 & 0.400 \\
 & Expanded & 19 & 6 & 6 & 0 & 1.000 & 0.316 \\
\addlinespace
APPALACHIA & Primary & 3 & 3 & 2 & 1 & 0.667 & 0.667 \\
 & Expanded & 5 & 3 & 3 & 0 & 1.000 & 0.600 \\
\bottomrule
\end{tabular}
\end{table}

\subsection{Outlier detection results}
Figures~\ref{fig:cars_outliers} and
\ref{fig:bushfire_outliers}, together with the
additional figures in \ref{app:outlier_results},
show the posterior distributions of the observation-specific
precision weights obtained with the CRM and the posterior
contamination probabilities obtained with the BRM. 
Overall, both models show very good agreement with the
outlier classifications reported in the literature,
with nearly perfect outlier rankings for several datasets.
The quantitative results in Table~\ref{tab:outlier_ranking}
confirm that both models provide accurate outlier rankings
for most datasets. In particular, for the CARS dataset, both models perfectly separate the 15 benchmark outliers from the remaining observations. For the MILK and APPALACHIA datasets, both models also achieve very high AP and AUC values. For the PULP FIBER dataset, both models achieve
considerably better ranking performance when
evaluated against the expanded benchmark.
The most pronounced difference between the two robust
models is observed for the BUSHFIRE dataset.
While the CRM perfectly ranks the outliers in both
the primary and the expanded benchmarks above the
remaining observations, the BRM achieves considerably
lower ranking accuracy.
As can be seen in Figure~\ref{fig:bushfire_outliers},
the CRM assigns systematically lower precision weights
to the observations belonging to both outlier clusters,
whereas the BRM assigns relatively low contamination
probabilities even to the benchmark outliers.
This finding is consistent with the greater flexibility
of continuous scale mixing observed in the simulation
study in Section~\ref{sec:experiments}, particularly
under extremely heavy tails.
Continuous precision weights allow for a more gradual
adjustment of the influence of individual observations
and may therefore be advantageous when observations
differ in their degree of atypicality. \\ 
Table~\ref{tab:outlier_threshold} shows the BRM classification
results using a posterior contamination probability threshold
of 0.5.
Good ranking performance does not necessarily imply good
classification performance at this threshold.
For the MILK dataset, only six of the 15 primary benchmark
outliers are identified, despite an AUC of 0.998.
For BUSHFIRE, no observations are classified as contaminated,
whereas for CARS, all 15 benchmark outliers are correctly
classified.
For PULP FIBER, the BRM identifies all four primary benchmark
outliers and five additional observations, four of which
belong to the expanded benchmark.
These results illustrate that posterior contamination
probabilities can provide accurate outlier rankings even
when classification at a fixed threshold shows limited
agreement with literature-based benchmarks.

\begin{figure}[p!]
\centering
\includegraphics[width=\textwidth]{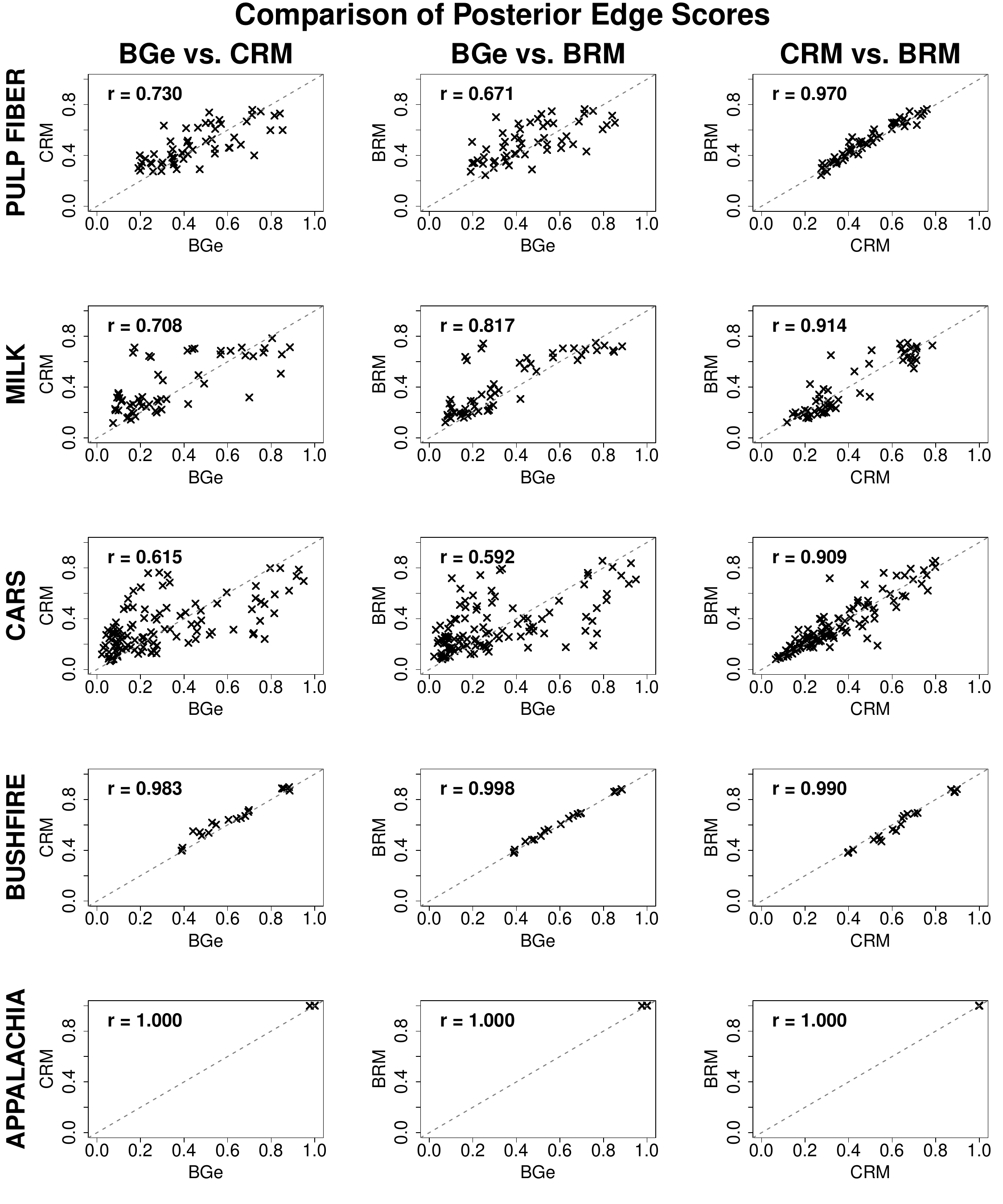}
\vspace{-0.75cm}
\caption{\textbf{Comparison of posterior edge probabilities obtained with the standard BGe model, the continuous robust model (CRM),
and the binary robust model (BRM) for five real-world datasets.}
Rows correspond to datasets, and columns to pairwise model
comparisons. Each cross represents a potential directed edge,
excluding self-loops. Posterior edge probabilities are computed
from sampled CPDAGs, with undirected edges contributing to both
directions. The dashed diagonal indicates identical edge
probabilities, while $r$ denotes the Pearson correlation
coefficient between the two sets of edge probabilities.}
\label{fig:edge_score_comparison}
\end{figure}

\subsection{Impact of robust modeling on network inference}

Figure~\ref{fig:edge_score_comparison} compares the
posterior edge probabilities obtained with the standard
BGe model and the two robust models.
For the PULP FIBER, MILK, and CARS datasets, robust
modeling has a substantial impact on the posterior
edge probabilities.
The largest differences are observed for the CARS dataset, for which the correlations between the posterior edge probabilities obtained with the standard BGe model and those obtained with the two robust models are both around 0.6.
Despite their different mixing mechanisms, the CRM
and the BRM yield highly correlated posterior edge
probabilities for all datasets, with correlation
coefficients exceeding 0.9.\\
Interestingly, for the BUSHFIRE dataset, all three
models yield very similar posterior edge probabilities,
despite the pronounced differences in outlier detection
performance between the two robust models.
This observation illustrates that the identification
and downweighting of atypical observations do not
necessarily lead to substantial changes in the
inferred dependence structure.
Similarly, for the APPALACHIA dataset, the posterior
edge probabilities are virtually identical across
the three models, although this result should be
interpreted in light of the small number of variables.
Overall, these results suggest that the impact of
robust modeling on network inference is dataset
dependent and that the presence of atypical observations
alone does not necessarily imply substantial changes
in the posterior support for individual edges.

\begin{figure}[t!]
\centering
\includegraphics[width=\textwidth]{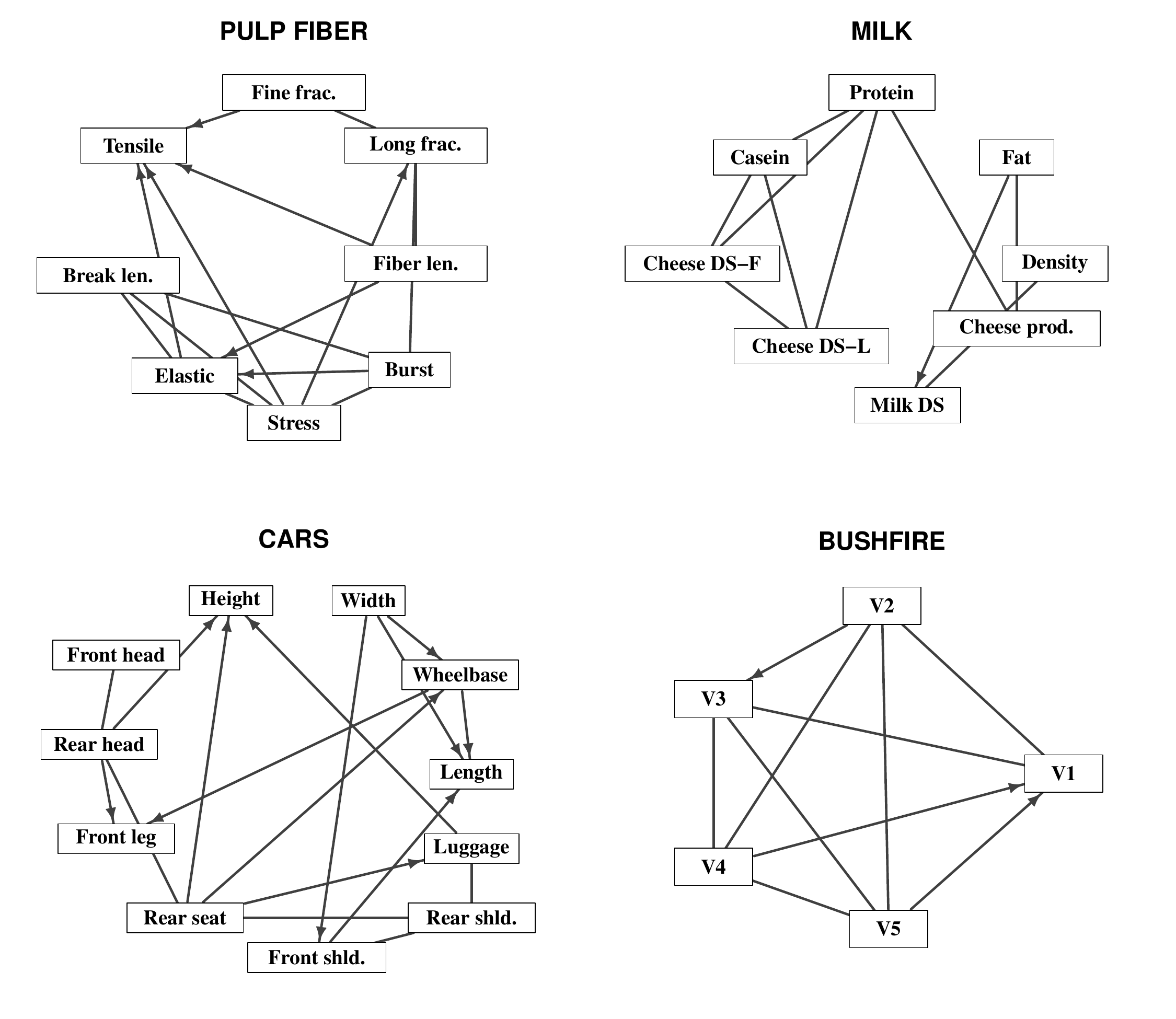}
\vspace{-1cm}
\caption{\textbf{Network structures inferred by the continuous robust
model (CRM) for four real-world datasets: PULP FIBER, MILK,
CARS, and BUSHFIRE.} Edges are included when their posterior
inclusion probabilities reach or exceed 0.5. Directed edges indicate connections for which only
one orientation reaches or exceeds the threshold,
whereas undirected edges indicate connections for
which both orientations reach or exceed the threshold. Posterior
edge probabilities are computed from sampled CPDAGs, with
undirected CPDAG edges contributing to both directions.
Node labels correspond to the variable abbreviations listed
in Table~\ref{tab:variable_names}. The APPALACHIA network
is omitted because it consists of only three nodes that
are pairwise connected by undirected edges.}
\label{fig:CRM_networks}
\end{figure}

To illustrate the network structures inferred with
the CRM, Figure~\ref{fig:CRM_networks} shows the
networks obtained by applying a threshold of 0.5
to the posterior edge probabilities for four of
the five real-world datasets.
The extracted networks contain both directed and
undirected edges, reflecting the posterior support
for connections between variables.
These networks provide compact summaries of the
inferred dependence structures after accounting
for atypical observations.
As discussed in Section~\ref{sec:ci}, the underlying
Gaussian DAG models describe conditional independence
relations conditional on the latent precision weights,
rather than conditional independence relations in the
marginal observation distribution.
These datasets are well established in the robust
statistics literature but appear to have received
little attention in Bayesian network structure learning.
Given their well-documented outlier classifications,
they may serve as useful benchmarks for future studies
on robust network structure learning.

\section{Discussion and conclusion}\label{sec:discussion}
In this paper, we have developed a unified framework
for robust Gaussian Bayesian network structure learning
based on Gaussian scale mixtures.
The main methodological contribution is a weighted BGe
score that is analytically tractable, decomposable, and
score equivalent conditional on the latent precision weights.
Score equivalence is also preserved after integrating out
these weights under a DAG-independent mixing prior.
The framework combines continuous and binary scale mixing
using the same network scoring mechanism and allows for
joint network structure learning and outlier detection.
However, unlike score equivalence, the conditional
independencies implied by the underlying Gaussian DAG
are not necessarily preserved when integrating out the
latent precision weights.
The inferred network therefore describes the Gaussian
dependence structure conditional on these weights. \\
Our simulation studies demonstrate that robust modeling
can substantially improve network structure recovery
in the presence of atypical observations without
noticeably reducing accuracy on uncontaminated Gaussian
data. The results under additive contamination show
that these improvements also extend to settings where
the assumed scale mixture models are misspecified.
While both robust models perform well under binary
scale mixing and additive contamination, the CRM is
considerably more successful under extremely heavy tails.
These results suggest that the CRM may be preferable
when the nature and severity of the contamination are
unknown. \\
The real-world applications demonstrate that both robust
models can clearly distinguish atypical observations
from regular observations in most datasets, with outlier
rankings that agree very well with literature-based
classifications.
The CRM identifies atypical observations through low
posterior mean precision weights, whereas the BRM
assigns posterior contamination probabilities that
also allow for an explicit classification of observations.
However, good ranking performance does not necessarily
imply good classification performance at a fixed
probability threshold.
Accounting for atypical observations can substantially
change the posterior edge probabilities, although the
extent of these changes is dataset dependent. \\
The framework has several limitations that suggest
directions for future research.
First, it targets casewise rather than cellwise
contamination. Extensions to cellwise contamination
would generally require variable-specific precision
weights or explicit measurement error models and would
not immediately inherit the same score equivalence
properties.
Second, Gaussian scale mixtures assume a common location
vector and proportional covariance matrices across
mixture components, thereby restricting the range of
contamination mechanisms.
Finally, although the weighted BGe score is decomposable
conditional on the latent precision weights,
updates of the latent precision weights generally require
recomputation of all local network scores.
More efficient inference algorithms would therefore be
valuable for applications to larger networks and datasets. \\
In conclusion, the proposed weighted BGe score extends
fully Bayesian Gaussian network structure learning to
data containing atypical observations and provides a
basis for incorporating robustness into other
extensions of the BGe framework.
\section*{Code availability}
Upon acceptance, the R code implementing the proposed
methods will be made publicly available through
a GitHub repository.



\clearpage
\newpage

\appendix
\section{Derivation of the weighted complete-DAG marginal likelihood}
\label{app:weighted}
We derive Eq.~\eqref{eq:weighted-subset} for a generic
$l$-dimensional variable subset $L$. Let
$\mathbf{W}_L=(\Bsigma^{L,L})^{-1}$ denote the precision matrix of
the corresponding marginal Gaussian distribution. The induced
Wishart prior is
$\mathbf{W}_L\sim\mathcal{W}_l(\alpha_{w,L},\Tdagger^{L,L})$,
where $\alpha_{w,L}=\alpha_w-n+l$. The corresponding conditional prior for the mean
vector is
\[
\Bmu^L\mid\mathbf{W}_L
\sim\mathcal{N}_l
\left(
\Bmu_0^L,(\alpha_\mu\mathbf{W}_L)^{-1}
\right).
\] The derivation is analogous to that of the standard BGe score, but is based on the weighted sufficient statistics.
Conditional on $\Blambda$, the likelihood is
\begin{align*}
 & (2\pi)^{-lm/2}
 |\mathbf{W}_L|^{m/2}
 \left(\prod_{i=1}^m\lambda_i\right)^{l/2}
\\[-1mm]
 & \quad\times \exp\!\left\{-\frac12\operatorname{tr}\left[
 \left(\mathbf{S}_{[\lambda]}^{L,L}
 +m_\lambda(\bar{\mathbf{x}}_{[\lambda]}^{L}-\Bmu^{L})
 (\bar{\mathbf{x}}_{[\lambda]}^{L}-\Bmu^{L})^{\transp}
 \right)\mathbf{W}_L\right]\right\}.
\end{align*}
Multiplying by the Gaussian prior on $\Bmu^L$ and completing the square
yields the posterior mean
\[
 \Bmu_{[\lambda]}^{\star,L}
 =\frac{\alpha_\mu\Bmu_0^L+m_\lambda
 \bar{\mathbf{x}}_{[\lambda]}^L}{\alpha_\mu+m_\lambda}
\]
and the quadratic term
\begin{align*}
 &\mathbf{S}_{[\lambda]}^{L,L}
 +\frac{\alpha_\mu m_\lambda}{\alpha_\mu+m_\lambda}
 (\bar{\mathbf{x}}_{[\lambda]}^L-\Bmu_0^L)
 (\bar{\mathbf{x}}_{[\lambda]}^L-\Bmu_0^L)^{\transp}.
\end{align*}
The resulting conditional posterior distribution
of the mean vector is
\[
\Bmu^L\mid(\mathbf{W}_L,\mathcal{D}^L,\Blambda)
\sim\mathcal{N}_l
\left(
\Bmu_{[\lambda]}^{\star,L},
((\alpha_\mu+m_\lambda)\mathbf{W}_L)^{-1}
\right).
\]
Integrating over $\Bmu^L$ produces the factor
$(\alpha_\mu/(\alpha_\mu+m_\lambda))^{l/2}$.
The posterior distribution of the precision matrix is Wishart,
\begin{equation}
 \mathbf{W}_L\mid(\mathcal{D}^L,\Blambda,\mathcal{G}_C^L)
 \sim\mathcal{W}_l
 (\alpha_{w,L}+m,\mathbf{T}_{\ddagger}^{L,L}).
\end{equation}
Integrating this Wishart kernel and retaining its normalizing constants
gives Eq.~\eqref{eq:weighted-subset}. In particular, the $2^{lm/2}$
ratio arising from the Wishart integral cancels the corresponding factor
in $(2\pi)^{-lm/2}$, leaving $\pi^{-lm/2}$.

\section{Proof of score equivalence}
\label{app:equiv}
For any fixed positive precision weight vector $\Blambda$,
the Gaussian likelihood continues to factorize according
to the DAG. As shown in
Eq.~\eqref{eq:weighted-local-regression}, the precision
weights affect only the conditional residual variances,
while the intercepts and regression coefficients remain
unchanged. Hence, the parameterization of the underlying
Gaussian Bayesian network remains unchanged.\\

We retain the prior construction underlying the classical
BGe score \citep{GeigerHeckerman2002}, whereby the local
parameter priors are consistently induced from a common
Normal--Wishart prior specified for a complete DAG.
In particular, global parameter independence and prior
modularity are preserved, since the precision weights
do not enter the prior distribution of the underlying
Gaussian parameters.
Moreover, complete DAGs remain equivalent
parameterizations of the same weighted Gaussian
likelihood for any fixed positive precision weight
vector.\\

The likelihood factorization, together with global
parameter independence, allows us to integrate over
the local parameter sets separately. Prior modularity
and the consistency of the induced Normal--Wishart
priors across variable subsets then allow us to express
the integrated local likelihoods as ratios of weighted
complete-DAG marginal likelihoods.
For any variable subset $L$, define the weighted
complete-DAG marginal likelihood
\begin{equation*}
 Q_{\Blambda}(L)
 :=
 p(\mathcal{D}^{L}\mid\Blambda,\mathcal{G}_C^{L}),
\end{equation*}
with $Q_{\Blambda}(\varnothing)=1$.
Consider a complete DAG on the variable set
$L\cup\{X_j\}$ in which $X_j$ is the last node in
the topological ordering. By global parameter
independence, the integrated likelihood factorizes
into the marginal likelihood for the variables in $L$
and the integrated local likelihood for $X_j$ given $L$.
Consequently, the latter can be expressed as
\begin{equation*}
 \frac{Q_{\Blambda}(L\cup\{X_j\})}
      {Q_{\Blambda}(L)}.
\end{equation*}
By prior modularity, the same local parameter prior
and hence the same local score apply to any DAG
in which $L$ is the parent set of $X_j$.\\

Thus, the weighted BGe score retains the factorization
\begin{equation*}
 p(\mathcal{D}\mid\mathcal{G},\Blambda)
 =
 \prod_{j=1}^n
 \frac{Q_{\Blambda}(\BPi_j\cup\{X_j\})}
      {Q_{\Blambda}(\BPi_j)}.
\end{equation*}

Importantly, $Q_{\Blambda}(L)$ depends only on the
variable subset $L$, the observed data, the common
prior hyperparameters, and the precision weights.
It does not depend on the orientation of the edges
in the complete DAG on $L$.\\

We now establish score equivalence using the fact
that two Markov-equivalent DAGs can be connected
by a sequence of covered edge reversals
\citep{Chickering1995}.
Consider a covered edge $X_a\rightarrow X_b$ and let
$L=\BPi_a=\BPi_b\setminus\{X_a\}$.
The product of the two affected local scores before
reversal is
\begin{align*}
 \frac{Q_{\Blambda}(L\cup\{X_a\})}
       {Q_{\Blambda}(L)}
 \cdot
 \frac{Q_{\Blambda}(L\cup\{X_a,X_b\})}
       {Q_{\Blambda}(L\cup\{X_a\})}
 =
 \frac{Q_{\Blambda}(L\cup\{X_a,X_b\})}
      {Q_{\Blambda}(L)}.
\end{align*}

After reversing the edge, the corresponding factors
become
\begin{align*}
 \frac{Q_{\Blambda}(L\cup\{X_b\})}
       {Q_{\Blambda}(L)}
 \cdot
 \frac{Q_{\Blambda}(L\cup\{X_a,X_b\})}
       {Q_{\Blambda}(L\cup\{X_b\})}
=
 \frac{Q_{\Blambda}(L\cup\{X_a,X_b\})}
      {Q_{\Blambda}(L)}.
\end{align*}

All other local factors are unchanged. Therefore,
the weighted BGe scores agree for a covered edge
reversal, and hence for any two Markov-equivalent
DAGs.
The argument holds for every fixed positive
$\Blambda$, independently of the mixing distribution.
\newpage
\section{Variable names and abbreviations.}
\label{app:variables}
\footnotesize
\begin{longtable}{@{}p{0.18\textwidth}p{0.55\textwidth}p{0.20\textwidth}@{}}
\caption{Variable names and abbreviations.}
\label{tab:variable_names}\\
\toprule
Dataset & Variable name & Abbreviation \\
\midrule
\endfirsthead

\toprule
Dataset & Variable name & Abbreviation \\
\midrule
\endhead

\midrule
\multicolumn{3}{r}{\emph{Continued on next page}}\\
\endfoot

\bottomrule
\endlastfoot

\multicolumn{3}{l}{\textbf{CARS}}\\
 & Length & Length \\
 & Wheelbase & Wheelbase \\
 & Width & Width \\
 & Height & Height \\
 & Front headroom & Front head. \\
 & Rear headroom & Rear head. \\
 & Front legroom & Front leg. \\
 & Rear seating room & Rear seat. \\
 & Front shoulder room & Front shld. \\
 & Rear shoulder room & Rear shld. \\
 & Luggage space & Luggage \\
\addlinespace

\multicolumn{3}{l}{\textbf{BUSHFIRE}}\\
 & Satellite bands 1-5 & V1-V5 \\
\addlinespace

\multicolumn{3}{l}{\textbf{PULP FIBER}}\\
 & Arithmetic fiber length & Fiber len. \\
 & Long fiber fraction & Long frac. \\
 & Fine fiber fraction & Fine frac. \\
 & Zero-span tensile & Tensile \\
 & Breaking length & Break len. \\
 & Elastic modulus & Elastic  \\
 & Stress at failure & Stress  \\
 & Burst strength & Burst  \\
\addlinespace

\multicolumn{3}{l}{\textbf{MILK}}\\
 & Density & Density \\
 & Fat content & Fat \\
 & Protein content & Protein \\
 & Casein content & Casein \\
 & Cheese dry substance (factory) & Cheese DS-F \\
 & Cheese dry substance (laboratory) & Cheese DS-L \\
 & Milk dry substance & Milk DS \\
 & Cheese product & Cheese prod. \\
\addlinespace

\multicolumn{3}{l}{\textbf{APPALACHIA}}\\
 & L-coefficient of variation & --- \\
 & L-coefficient of skewness & --- \\
 & L-coefficient of kurtosis & --- \\
\end{longtable}
\normalsize

\clearpage
\section{Additional outlier detection results}
\label{app:outlier_results}

This appendix complements the real-world data analysis
in Section~\ref{sec:real_data} by providing detailed
descriptions of the literature-based outlier benchmarks
and the corresponding outlier detection figures for the
PULP FIBER, MILK, and APPALACHIA datasets.\\

\begin{figure}[t!]
\centering
\includegraphics[width=\textwidth]{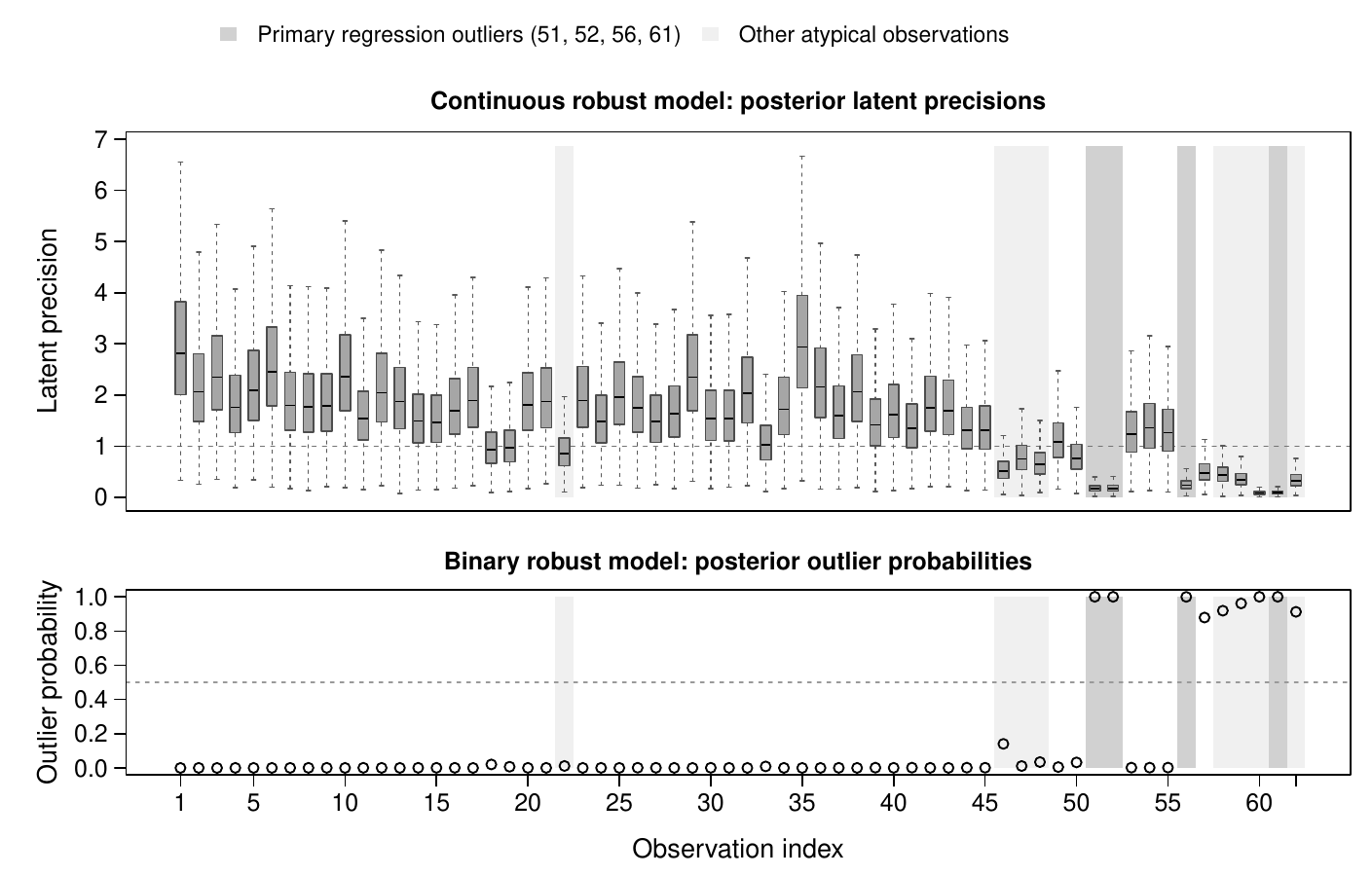}
\caption{\textbf{Outlier detection in the PULP FIBER dataset.}
Dark gray shading indicates the primary regression outliers,
while light gray shading indicates additional atypical
observations.
The visualization conventions are as in
Figure~\ref{fig:cars_outliers}.}
\label{fig:pulp_outliers}
\end{figure}

The PULP FIBER dataset consists of 62 observations and has
been investigated in several studies on robust multivariate
regression and outlier detection
\citep{Rousseeuw2004,OngAlih2015}. In particular, \citet{Rousseeuw2004} identified four
observations (51, 52, 56, and 61) as outliers using
univariate least trimmed squares regression.
Subsequently, \citet{OngAlih2015} identified 12
out-of-control observations using a multivariate
control-chart approach.
For visualization, we distinguish two groups.
The first comprises the four observations identified
by \citet{Rousseeuw2004}, while the second consists
of the eight additional observations
(22, 46--48, 58--60, and 62).
These groups are highlighted using different shades
of gray in Figure~\ref{fig:pulp_outliers}.\\

\begin{figure}[t!]
\centering
\includegraphics[width=\textwidth]{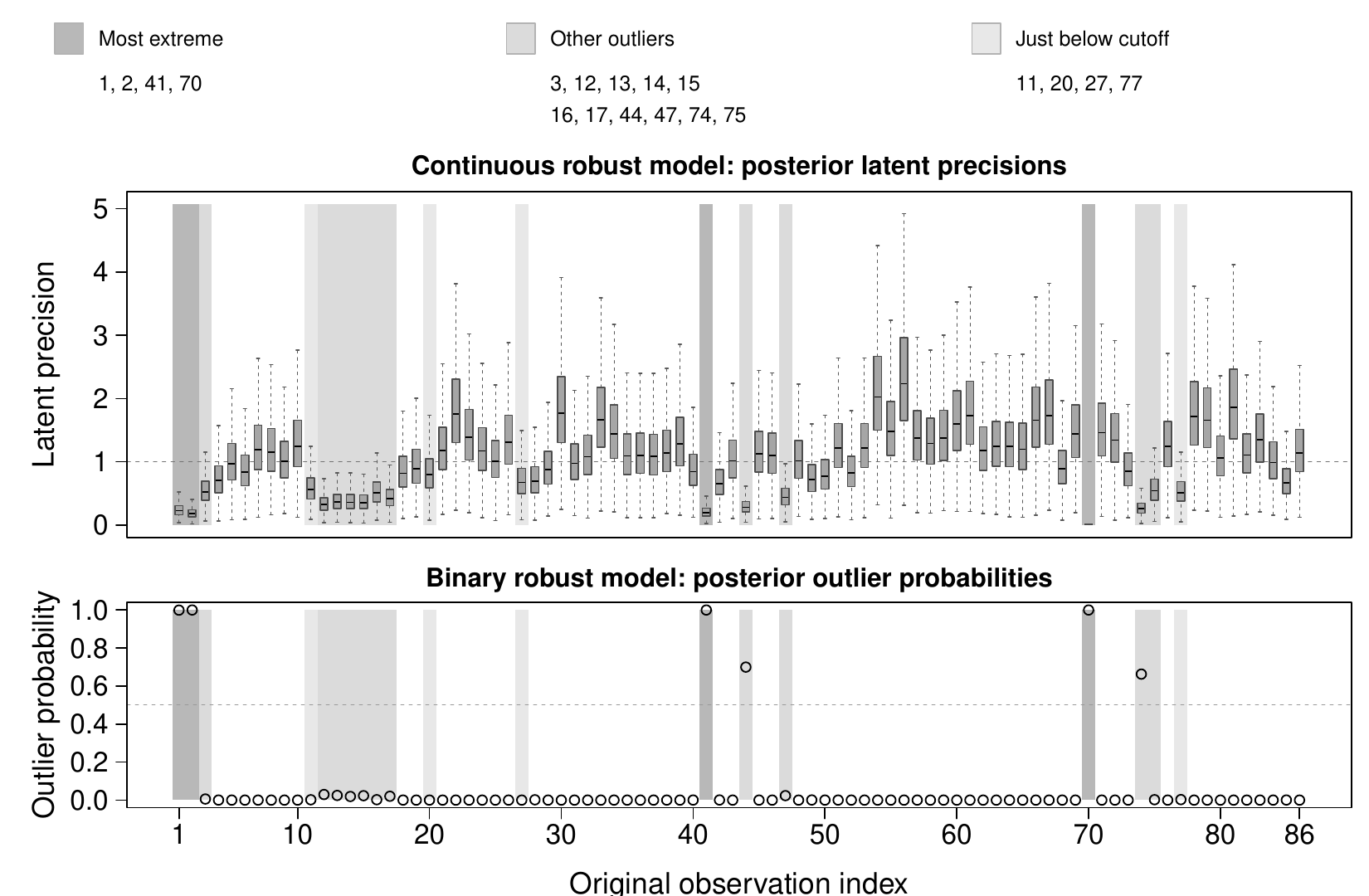}
\caption{\textbf{Outlier detection in the MILK dataset.}
Three shades of gray distinguish the most extreme outliers,
the remaining observations exceeding the cutoff, and the
observations just below the cutoff.
The visualization conventions are as in
Figure~\ref{fig:cars_outliers}.}
\label{fig:milk_outliers}
\end{figure}

The MILK dataset has been extensively studied in the literature on
multivariate outlier detection
\citep{Atkinson1994,RockeWoodruff1996}.
In particular, \citet{RockeWoodruff1996} investigated the data
using a hybrid robust estimation procedure based on Mahalanobis
distances and identified 15 observations exceeding a 99\% cutoff.
For the purpose of visualization, we distinguish three groups
of potentially outlying observations based on their findings.
The first group comprises the four most extreme observations
(1, 2, 41, and 70), which are clearly separated from the
remaining observations.
The second group consists of the other 11 observations exceeding
the cutoff (3, 12--17, 44, 47, 74, and 75).
Finally, the third group contains four additional observations
(11, 20, 27, and 77) that fall just below the cutoff and may
therefore also be considered potentially atypical.
These three groups are highlighted using different shades of gray
in Figure~\ref{fig:milk_outliers}, providing an external reference
against which the outlier detection results of our robust
Gaussian Bayesian network models can be compared.\\

\begin{figure}[t!]
\centering
\includegraphics[width=\textwidth]
{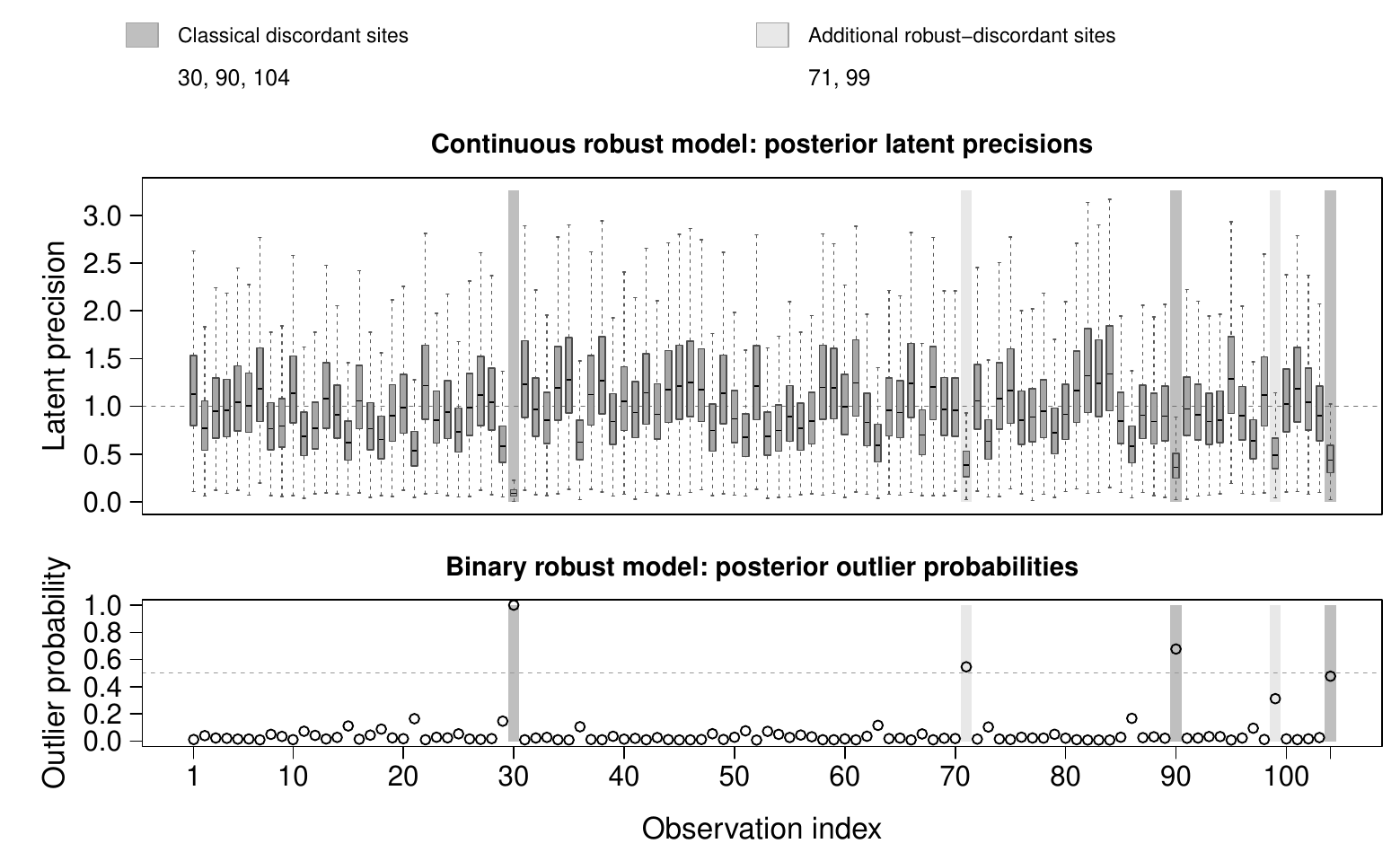}
\caption{\textbf{Outlier detection in the APPALACHIA dataset.}
Dark gray shading indicates observations identified by both
classical and robust discordancy measures, while light gray
shading indicates observations identified only by the
robust measure.
The visualization conventions are as in
Figure~\ref{fig:cars_outliers}.}
\label{fig:appalachia_outliers}
\end{figure}

The APPALACHIA dataset consists of 104 observations and three
variables describing annual maximum streamflow characteristics
at gauging stations in the central Appalachia region of the
United States. The variables are the L-moment ratios L-CV,
L-skewness, and L-kurtosis.
The dataset was investigated by \citet{Neykov2007} using
classical and robust discordancy measures.
Their analysis identified five potentially
outlying observations, which we divide into two groups.
The first group consists of observations 30, 90, and 104,
which were identified by both the classical and robust
discordancy measures.
The second group comprises observations 71 and 99,
which were additionally identified by the robust measure
based on the minimum covariance determinant estimator.
These two groups are highlighted using different shades
of gray in Figure~\ref{fig:appalachia_outliers},
providing an external benchmark for evaluating the
outlier detection capabilities of our robust Gaussian
Bayesian network models.

\clearpage

\scriptsize
\bibliographystyle{plainnat}
\bibliography{tdh}

@article{SCUTARI_NEERLANDICA,
author = {Scutari, M.},
title = {{{Bayesian}} network models for incomplete and dynamic data},
journal = {Statistica Neerlandica},
volume = {74},
issue = {3},
pages = {397-419},
year = {2020}
}

@article{GrzegorczykHusmeierBioinf08,
   author    = {Grzegorczyk, M. and Husmeier, D. and Edwards, K. and Ghazal, P. and Millar, A.},
   title     = {Modelling non-stationary gene regulatory processes with a non-homogeneous {{Bayesian}} network and the allocation sampler},
   journal   = {Bioinformatics},
   year      = {2008},
   volume    = {24},
   number    = {18},
   pages     = {2071--2078}
}

@article{SCUTARI_REASONING,
author = {Bodewes, T. and Scutari, M.},
journal = {International Journal of Approximate Reasoning},
title = {Learning {{Bayesian}} networks from incomplete data with the node-average likelihood},
year = {2021},
volume = {138},
issue = {3},
pages = {145-160}
}

@article{OSEM,
author = {Luo, X.G. and Moffa, G. and Kuipers, J.},
title = {Learning {{Bayesian}} Networks from Ordinal Data},
year = {2021},
journal = {Journal of Machine Learning Research},
volume = {22},
pages = {1-44}
}

@inproceedings{nikzad2025scalingbayesiandagsampling,
  author    = {Nikzad, D. and Zhilkin, A. and Harviainen, J.
               and Kuipers, J. and Moffa, G. and Koivisto, M.},
  title     = {Scaling Up {{Bayesian}} {{DAG}} Sampling},
  booktitle = {Proceedings of the 42nd Conference on
               Uncertainty in Artificial Intelligence},
  series    = {Proceedings of Machine Learning Research},
  volume    = {337},
  pages     = {4949--4971},
  publisher = {PMLR},
  year      = {2026}
}

@article{Kuipers_2017,
author = {Kuipers, J. and Moffa, G.},
year = {2017},
volume= {112},
issue = {517},
pages = {282-299},
title = {Partition {{MCMC}} for Inference on Acyclic Digraphs},
journal = {Journal of the American Statistical Association},
}

@article{GrzegorczykHusmeierMach08,
   author    = {Grzegorczyk, M. and Husmeier, D.},
   title     = {{Improving the structure MCMC sampler for {{Bayesian}} networks by introducing a new edge reversal move}},
   journal   = {Machine Learning},
   year      = {2008},
   volume    = {71},
   number    = {2-3},
   pages     = {265--305}
}

@article{FriedKollerOrderMCMC,
   author    = {Friedman, N. and Koller, D.},
   title     = {Being {{Bayesian}} about Network Structure: A {{Bayesian}} Approach to Structure Discovery in {{Bayesian}} Networks},
   journal   = {Machine Learning},
   year      = {2003},
   volume = {50},
   pages = {95-126}
}

@article{KALISCH_ROBUST,
  author  = {Kalisch, M. and B{\"u}hlmann, P.},
  title   = {Robustification of the {PC}-algorithm for
             directed acyclic graphs},
  journal = {Journal of Computational and Graphical Statistics},
  year    = {2008},
  volume  = {17},
  number  = {4},
  pages   = {773--789},
  doi     = {10.1198/106186008X381927}
}

@inproceedings{VermaPearl1990,
  author    = {Verma, T. and Pearl, J.},
  title     = {Equivalence and Synthesis of Causal Models},
  booktitle = {Proceedings of the Sixth Conference on
               Uncertainty in Artificial Intelligence},
  pages     = {255--270},
  year      = {1990},
  publisher = {Elsevier}
}

@inproceedings{Scutari2022,
  title     = {Using Mixed-Effects Models to Learn {{Bayesian}} Networks from Related Data Sets},
  author    = {Scutari, M. and Marquis, C. and Azzimonti, L.},
  booktitle = {International Conference on Probabilistic Graphical Models},
  series    = {Proceedings of Machine Learning Research},
  volume    = {186},
  pages     = {73--84},
  year      = {2022},
}

@article{Kitson2023,
  title   = {A survey of {{Bayesian}} Network structure learning},
  author  = {Kitson, N. K. and Constantinou, A. C. and Guo, Z. and Liu, Y. and Chobtham, K.},
  journal = {Artificial Intelligence Review},
  year    = {2023},
  volume  = {56},
  pages   = {8721--8814}
}

@article{FriedmanEtAl2000,
  author  = {Friedman, N. and Linial, M. and Nachman, I. and Pe'er, D.},
  title   = {Using {{Bayesian}} networks to analyze expression data},
  journal = {Journal of Computational Biology},
  year    = {2000},
  volume  = {7},
  number  = {3--4},
  pages   = {601--620},
  doi     = {10.1089/106652700750050961}
}

@book{KollerFriedman2009,
  author    = {Koller, D. and Friedman, N.},
  title     = {Probabilistic Graphical Models:
               Principles and Techniques},
  publisher = {MIT Press},
  address   = {Cambridge, MA, USA},
  year      = {2009}
}

@book{ScutariDenis2014,
  author    = {Scutari, M. and Denis, J.-B.},
  title     = {{Bayesian} Networks: With Examples in {R}},
  publisher = {Chapman \& Hall},
  address   = {Boca Raton, FL, USA},
  year      = {2014}
}

@article{KuipersEtAl2022,
  author  = {Kuipers, J. and Suter, P. and Moffa, G.},
  title   = {Efficient Sampling and Structure Learning
             of {{Bayesian}} Networks},
  journal = {Journal of Computational and Graphical Statistics},
  year    = {2022},
  volume  = {31},
  number  = {3},
  pages   = {639--650},
  doi     = {10.1080/10618600.2021.2020127}
}

@article{SOFTWARE3,
 title={Bayesian Structure Learning and Sampling of {{Bayesian}} Networks with the {{R}} Package {{BiDAG}}},
 volume={105},
 number={9},
 journal={Journal of Statistical Software},
 author={Suter, P. and Kuipers, J. and Moffa, G. and Beerenwinkel, N.},
 year={2023},
 pages={1–31}
}

@article{Chickering2002,
  author  = {Chickering, D.M.},
  title   = {Optimal Structure Identification with Greedy Search},
  journal = {Journal of Machine Learning Research},
  year    = {2002},
  volume  = {3},
  pages   = {507--554}
}

@article{TsamardinosEtAl2006,
  author  = {Tsamardinos, I. and Brown, L.E. and Aliferis, C.F.},
  title   = {The Max-Min Hill-Climbing {{Bayesian}}
             Network Structure Learning Algorithm},
  journal = {Machine Learning},
  year    = {2006},
  volume  = {65},
  number  = {1},
  pages   = {31--78},
  doi     = {10.1007/s10994-006-6889-7}
}

@article{Scutari2010,
  author  = {Scutari, M.},
  title   = {Learning {{Bayesian}} Networks with the
             {bnlearn} {R} Package},
  journal = {Journal of Statistical Software},
  year    = {2010},
  volume  = {35},
  number  = {3},
  pages   = {1--22},
  doi     = {10.18637/jss.v035.i03}
}

@InProceedings{PGM2026,
  title = 	 {Hierarchically Coupled {{Gaussian}} {{Bayesian}} Networks},
  author =       {Ma, C. and Grzegorczyk, M.},
  booktitle = 	 {Proceedings of the 13th International Conference on Probabilistic Graphical Models},
  pages = 	 {523--544},
  year = 	 {2026},
  editor = 	 {Camps-Valls, G. and Leonelli, M. and Varando, G.},
  volume = 	 {343},
  series = 	 {Proceedings of Machine Learning Research},
  publisher =    {PMLR},
}

@article{IJAR_Kezhuo,
title = {Two {{Bayesian}} approaches to dynamic {{Gaussian}} {{Bayesian}} networks with intra- and inter-slice edges},
journal = {International Journal of Approximate Reasoning},
volume = {201},
note = {Article 109860},
year = {2027},
author = {Li, K. and Grzegorczyk, M.},
}

@article{grzegorczyk_TIED,
  title = {A {{BGe}} score for tied-covariance mixture
           {{Gaussian}} {{Bayesian}} networks},
  author = {Grzegorczyk, M.},
  journal = {International Journal of Approximate Reasoning},
  year = {2026},
  volume = {198},
  note = {Article 109771},
}

@article{SaitoRehmsmeier2017,
  title = {Precrec: fast and accurate precision-recall
           and {ROC} curve calculations in {R}},
  author = {Saito, T. and Rehmsmeier, M.},
  journal = {Bioinformatics},
  year = {2017},
  volume = {33},
  number = {1},
  pages = {145--147},
}

@article{FinegoldDrton2011,
  author  = {Finegold, M. and Drton, M.},
  title   = {Robust graphical modeling of gene networks
             using classical and alternative {$t$}-distributions},
  journal = {The Annals of Applied Statistics},
  year    = {2011},
  volume  = {5},
  number  = {2A},
  pages   = {1057--1080},
}

@article{FinegoldDrton2014,
  author  = {Finegold, M. and Drton, M.},
  title   = {Robust {{Bayesian}} graphical modeling using
             {{Dirichlet}} {$t$}-distributions},
  journal = {Bayesian Analysis},
  year    = {2014},
  volume  = {9},
  number  = {3},
  pages   = {521--550},
}

@article{ChakrabortyEtAl2024,
  author  = {Chakraborty, M. and Baladandayuthapani, V. and Bhadra, A. and Ha, M.J.},
  title   = {Bayesian robust learning in chain graph
             models for integrative pharmacogenomics},
  journal = {The Annals of Applied Statistics},
  year    = {2024},
  volume  = {18},
  number  = {4},
  pages   = {3274--3296},
}

@article{HarrisDrton2013,
  author  = {Harris, N. and Drton, M.},
  title   = {{PC} algorithm for nonparanormal
             graphical models},
  journal = {Journal of Machine Learning Research},
  year    = {2013},
  volume  = {14},
  pages   = {3365--3383}
}

@misc{NazariEtAl2026,
  author  = {Nazari, S. and Arashi, M. and Sadeghkhani, A.},
  title   = {Bayesian {{DAG}} structure learning with
             simultaneous shrinkage covariance
             estimation under scale-mixture error
             distributions in the proportional
             high-dimensional regime},
  year    = {2026},
  note    = {arXiv preprint arXiv:2607.08428},
}

@article{KuipersEtAl2014,
  author  = {Kuipers, J. and Moffa, G. and Heckerman, D.},
  title   = {Addendum on the scoring of {{Gaussian}}
             directed acyclic graphical models},
  journal = {The Annals of Statistics},
  year    = {2014},
  volume  = {42},
  number  = {4},
  pages   = {1689--1691},
}

@article{Neykov2007,
  author  = {Neykov, N.M. and Neytchev, P.N. and Van Gelder, P.H.A.J.M. and Todorov, V.K.},
  title   = {Robust Detection of Discordant Sites
             in Regional Frequency Analysis},
  journal = {Water Resources Research},
  year    = {2007},
  volume  = {43},
  number  = {6},
  pages   = {W06417},
}

@article{MaronnaYohai1995,
  author  = {Maronna, R.A. and Yohai, V.J.},
  title   = {The Behavior of the {Stahel-Donoho}
             Robust Multivariate Estimator},
  journal = {Journal of the American Statistical Association},
  year    = {1995},
  volume  = {90},
  number  = {429},
  pages   = {330--341},
}

@article{BeguinHulliger2008,
  author  = {B{\'e}guin, C. and Hulliger, B.},
  title   = {The {BACON-EEM} Algorithm for Multivariate
             Outlier Detection in Incomplete Survey Data},
  journal = {Survey Methodology},
  year    = {2008},
  volume  = {34},
  number  = {1},
  pages   = {91--103}
}

@article{Hubert2005,
  author  = {Hubert, M. and Rousseeuw, P.J. and Vanden Branden, K.},
  title   = {{ROBPCA}: A New Approach to Robust
             Principal Component Analysis},
  journal = {Technometrics},
  year    = {2005},
  volume  = {47},
  number  = {1},
  pages   = {64--79},
}

@article{OngAlih2015,
  author  = {Ong, H.C. and Alih, E.},
  title   = {A Control Chart Based on Cluster-Regression
             Adjustment for Retrospective Monitoring of
             Individual Characteristics},
  journal = {PLOS ONE},
  year    = {2015},
  volume  = {10},
  number  = {4},
  pages   = {e0125835},
}

@article{AndrewsMallows1974,
  author  = {Andrews, D.F. and Mallows, C.L.},
  title   = {Scale Mixtures of Normal Distributions},
  journal = {Journal of the Royal Statistical Society: Series B},
  year    = {1974},
  volume  = {36},
  number  = {1},
  pages   = {99--102},
}

@inproceedings{Chickering1995,
  author    = {Chickering, D.M.},
  title     = {A Transformational Characterization of Equivalent
               {{Bayesian}} Network Structures},
  booktitle = {Proceedings of the Eleventh Conference on
               Uncertainty in Artificial Intelligence},
  year      = {1995},
  pages     = {87--98}
}

@article{GeigerHeckerman2002,
  author  = {Geiger, D. and Heckerman, D.},
  title   = {Parameter Priors for Directed Acyclic Graphical Models
             and the Characterization of Several Probability Distributions},
  journal = {The Annals of Statistics},
  year    = {2002},
  volume  = {30},
  number  = {5},
  pages   = {1412--1440},
}

@article{LangeLittleTaylor1989,
  author  = {Lange, K.L. and Little, R.J.A. and Taylor, J.M.G.},
  title   = {Robust Statistical Modeling Using the {$t$} Distribution},
  journal = {Journal of the American Statistical Association},
  year    = {1989},
  volume  = {84},
  number  = {408},
  pages   = {881--896},
}

@article{PunzoMcNicholas2016,
  author  = {Punzo, A. and McNicholas, P.D.},
  title   = {Parsimonious Mixtures of Multivariate Contaminated
             Normal Distributions},
  journal = {Biometrical Journal},
  year    = {2016},
  volume  = {58},
  number  = {6},
  pages   = {1506--1537},
}

@article{Atkinson1994,
  author  = {Atkinson, A.C.},
  title   = {Fast Very Robust Methods for the Detection
             of Multiple Outliers},
  journal = {Journal of the American Statistical Association},
  year    = {1994},
  volume  = {89},
  number  = {428},
  pages   = {1329--1339},
}

@article{RockeWoodruff1996,
  author  = {Rocke, D.M. and Woodruff, D.L.},
  title   = {Identification of Outliers in Multivariate Data},
  journal = {Journal of the American Statistical Association},
  year    = {1996},
  volume  = {91},
  number  = {435},
  pages   = {1047--1061},
}

@article{Rousseeuw2004,
  author  = {Rousseeuw, P.J. and Van Aelst, S. and Van Driessen, K. and Agull{\'o}, J.},
  title   = {Robust Multivariate Regression},
  journal = {Technometrics},
  year    = {2004},
  volume  = {46},
  number  = {3},
  pages   = {293--305},
}

@article{Gryze2023,
author = {Grzegorczyk, M.}, 
year = {2023},
title = {Being {{Bayesian}} about learning {{Gaussian}} {{Bayesian}} networks from incomplete data},
journal = {International Journal of Approximate Reasoning},
volume = {160}, 
note = {Article 108954}
}

@article{Gryze2024,
author = {Grzegorczyk, M.}, 
year = {2024},
title = {Being {{Bayesian}} about learning  {{Bayesian}} networks from ordinal data},
journal = {International Journal of Approximate Reasoning},
volume = {170}, 
note = {Article 109205}
}

@article{Gryze2025,
author = {Grzegorczyk, M.}, 
year = {2025},
title = {Being {{Bayesian}} about learning  {{Bayesian}} networks from hybrid data},
journal = {International Journal of Approximate Reasoning},
volume = {187}, 
note = {Article 109549}
}

@inproceedings{GeigerHeckGaussUAI,
   author={Geiger, D. and Heckerman, D.},
   title={Learning {{Gaussian}} networks},
   year={1994},
   publisher={Morgan Kaufmann},
   address={San Francisco, CA.},
   pages={235-243},
   booktitle={Proceedings of the Tenth Conference on Uncertainty in Artificial Intelligence}
}
\end{document}